\documentclass[fleqn,usenatbib]{mnras}

\usepackage[T1]{fontenc}

\DeclareRobustCommand{\VAN}[3]{#2}
\let\VANthebibliography\thebibliography
\def\thebibliography{\DeclareRobustCommand{\VAN}[3]{##3}\VANthebibliography}

\usepackage{graphicx}	
\usepackage{amsmath}	
\usepackage{amssymb}	
\usepackage{newtxtext,newtxmath}

\title[No need for dark matter in the UDG AGC~114905]{No need for dark matter: resolved kinematics of the ultra-diffuse galaxy AGC 114905}

\author[Pavel E. Mancera Pi\~na et al.]{Pavel E. Mancera Pi\~na$^{1,2}$\thanks{e-mail: pavel@astro.rug.nl},
Filippo Fraternali$^{1}$, Tom Oosterloo$^{2,1}$, Elizabeth A K. Adams$^{2,1}$, \newauthor Kyle A. Oman$^{3}$, and Lukas Leisman$^{4,5}$\\
$^{1}$ Kapteyn Astronomical Institute, University of Groningen, Landleven 12, 9747 AD, Groningen, The Netherlands\\
$^{2}$ ASTRON, Netherlands Institute for Radio Astronomy, Oude Hoogeveensedijk 4, 7991 PD, Dwingeloo, The Netherlands\\
$^{3}$ Institute for Computational Cosmology, Department of Physics, Durham University, South Road, Durham DH1 3LE, United Kingdom\\
$^{4}$ Department of Physics and Astronomy, Valparaiso University, 1610 Campus Drive East, Valparaiso, IN 46383, USA\\
$^{5}$ Department of Astronomy, University of Illinois, 1002 W. Green St., Urbana, IL 61801, USA\\
}

\date{Accepted XXX. Received YYY; in original form ZZZ}

\pubyear{2015}

\begin{document}
\label{firstpage}
\pagerange{\pageref{firstpage}--\pageref{lastpage}}
\maketitle

\begin{abstract}
We present new H\,{\sc i} interferometric observations of the gas-rich ultra-diffuse galaxy AGC~114905, which previous work, based on low-resolution data, identified as an outlier of the baryonic Tully-Fisher relation. The new observations, at a spatial resolution $\sim 2.5$ times higher than before, reveal a regular H\,{\sc i} disc rotating at about 23~km~s$^{-1}$. Our kinematic parameters, recovered with a robust 3D kinematic modelling fitting technique, show that the flat part of the rotation curve is reached. Intriguingly, the rotation curve can be explained almost entirely by the baryonic mass distribution alone. We show that a standard cold dark matter halo that follows the concentration-halo mass relation fails to reproduce the amplitude of the rotation curve by a large margin. Only a halo with an extremely (and arguably unfeasible) low concentration reaches agreement with the data. We also find that the rotation curve of AGC~114905 deviates strongly from the predictions of Modified Newtonian dynamics. The inclination of the galaxy, which is measured independently from our modelling, remains the largest uncertainty in our analysis, but the associated errors are not large enough to reconcile the galaxy with the expectations of cold dark matter or Modified Newtonian dynamics.
\end{abstract}




\section{Introduction}

The properties, origin, and formation mechanisms of ultra-diffuse galaxies (UDGs) have been widely discussed in the last years. UDGs \citep{vandokkum2015} are low surface brightness galaxies (e.g. \citealt{impey1988}, see also discussion in \citealt{conselice2018}) with an extended light distribution. At fixed stellar mass or luminosity, UDGs have significantly larger effective radii than the `classical' dwarf galaxy population (e.g. \citealt{vandokkum2015,mihos2015, udgs_kiwics,chamba_udgs}).

UDGs are mostly found by number in massive galaxy clusters, but they are also present in galaxy groups, in the field, and even in voids (e.g. \citealt{vanderburg2016,leisman2017, roman_outsideclusters,udgs_kiwics,roman_udginvoid,barbosa_udgs2020,karunakaran2020}). The ubiquity of UDGs across different environments indicates that even if some of them form due to environmental processes, this is not the case for all UDGs, and they can also form due to their own internal processes. The population of UDGs is likely a mixed bag of galaxies with similar sizes and surface brightness, but perhaps multiple formation mechanisms.

The above idea seems confirmed by a number of results from semi-analytic models and hydrodynamics simulations that produce UDG-like simulated galaxies based on different physical processes. On the one hand, different authors report that classical dwarf galaxies can become UDGs (i.e. increase their size and likely decrease their surface brightness) due to cluster pre-processing phenomena such as tidal striping and tidal heating \citep{carleton_udgs,tremmel_udgs,sales_udgs}. On the other hand, it has also been suggested that internal processes can explain the optical properties of UDGs. \citet{amorisco2016} proposed a model where UDGs form due to a higher than average dark-matter angular momentum, which then gives rise to an extended stellar effective radius (see also \citealt{rong_udgs}). Here it is worth mentioning that even if UDGs inhabit normal dark matter haloes, they could still have a higher than average retained fraction of stellar specific angular momentum \citep{posti_spin,huds2020}. Another idea is that UDGs are dwarfs that became larger due to feedback-driven outflows, which change the dark matter and baryonic potential and allow the stars to migrate to more external orbits, increasing their effective radius \citep{nihao_udgs,fire_udgs}. It has also been argued that the expansion of the stellar orbits can be the result of massive mergers at early ($z > 1$) epochs  \citep{wright_udgs} or the by-product of very efficient globular cluster feedback \citep{trujillogomez_udgs}.

While the above models and simulations seem to produce simulated galaxies that match a number of properties of real UDGs, surprising observations of two different populations of UDGs have been more challenging to reproduce. First, it has been found that two gas-poor UDGs (DF--2 and DF--4) near (at least in projection) the galaxy NGC~1052, contain significantly less dark matter than expected based on stellar and globular cluster kinematics (e.g. \citealt{vandokkum_DF2,vandokkum_DF4,danieli_DF2,emsellem_DF2}). While caveats regarding the distance and accuracy of the kinematic tracers used to study these UDGs have been raised (e.g. \citealt{laporte_udgs,trujillo_distanceDF2}), DF--2 and DF--4 have motivated multiple studies aiming to explain the existence of dark-matter free galaxies. The main ideas to explain their existence involve dark matter removing mechanisms such as high-velocity collisions and tidal stripping, or a tidal dwarf origin (\citealt{haslbauer_UDGsnoDMIllustris,silk_udgs, montes_df4,shin_udgs,doppel_udgs,jackson_tidalUDGs}).

The second set of puzzling observations, still very much lacking a quantitative explanation, is related to the H\,{\sc i} kinematics of some gas-rich UDGs (sometimes also called H\,{\sc i}-bearing UDGs). Using unresolved ALFALFA data (see \citealt{alfalfa}), \citet[][see also \citealt{jones_huds,karunakaran2020}]{leisman2017} first found that gas-rich UDGs have narrow global H\,{\sc i} profiles for their gas mass. Then, \citet{huds2019,huds2020} studied a set of six of those gas-rich UDGs with low-resolution (two independent resolution elements per galaxy side) interferometric H\,{\sc i} observations. Using a state-of-the-art kinematic modelling fitting technique ($\mathrm{^{3D}}$Barolo, \citealt{barolo}) to overcome beam smearing effects, they recovered the circular speeds of their galaxies, unveiling two intriguing features. First, that having a baryonic mass a factor $10-100$ larger than galaxies with similar circular speed, H\,{\sc i}-rich UDGs shift off the baryonic Tully-Fisher relation (BTFR, \citealt{mcgaugh2000}), with the offset from the BTFR correlating with the UDGs optical disc scale lengths. And second, that their dynamical mass within the extent of the H\,{\sc i} disc is about the same as their baryonic mass, meaning that the galaxies have very low dark matter fractions within scales as large as 10~kpc. These features suggest that gas-rich UDGs have atypical non-luminous mass distributions, making them a promising population to test dark matter theories. It is also important to stress that these gas-rich UDGs are selected to be fairly isolated \citep{leisman2017}, and they lie at distances of several tens of Mpc where Hubble flow distances are robust, negating some of the main concerns raised for DF--2 and DF--4.

Given all this, it is important to further characterise the properties of these UDGs, which are apparently gas-rich but also dark-matter poor (at least within the observed radii). One way to do this is by studying their H\,{\sc i} rotation curves, as H\,{\sc i} provides arguably the best kinematic tracer for disc galaxies, both massive and dwarfs (e.g. \citealt{begeman,deblok08,iorio}). 

In this work, we present and analyse new, high-resolution interferometric observations of one of these peculiar gas-rich UDGs, AGC~114905. As we will show, the galaxy seems to pose a challenge to the currently favoured galaxy formation framework. This paper is organised in the following way. In Section~ \ref{sec:data}, we describe the main properties of AGC~114905 and we present the data used in this work. In Section~ \ref{sec:kinematics} we show the kinematic modelling of the galaxy, and in Section~ \ref{sec:massmodels} the resulting mass models. We discuss our results in Section~ \ref{sec:discussion}, to then present our conclusions and summary in Section~ \ref{sec:conclusions}.

\section{Data and properties of AGC 114905}
\label{sec:data}
\begin{figure*}
    \centering
    \includegraphics[scale=0.514]{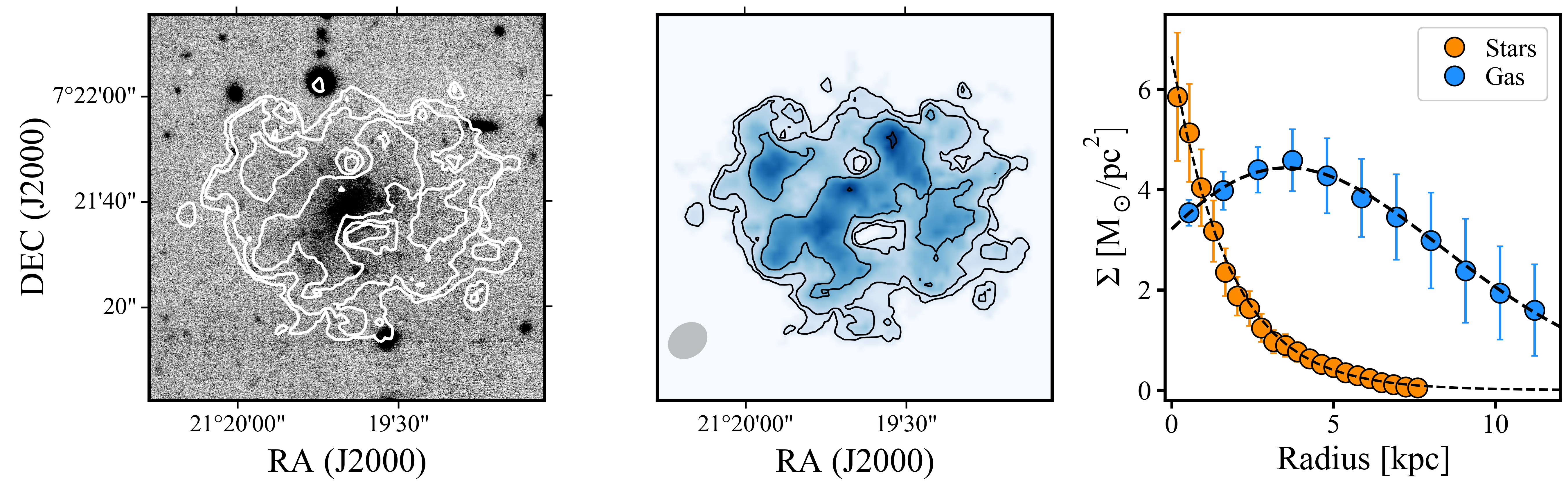}
    \caption{\textit{Left:} Stellar image of AGC~114905 with the total H\,{\sc i} contours overlaid. The contours are at 1, 2, 4$\times 10^{20}$~atoms cm$^{-2}$, the noise level is 4.1$\times 10^{19}$ atoms~cm$^{-2}$. \textit{Middle:} Total H\,{\sc i} intensity map; contours as in the previous panel. The grey ellipse shows the beam of our data. \textit{Right:} Stellar (orange) and gas (blue, includes helium correction) surface mass density profiles of AGC~114905. The dashed black lines on top show the fits to the distributions used to obtain the stellar and gas circular speeds (see Section~ \ref{sec:massmodels}).}
    \label{fig:sds}
\end{figure*}

AGC~114905 (01:25:18.60, +07:21:41.11, J2000) lies at a (Hubble-flow) distance $D = 76 \pm 5$~Mpc \citep{leisman2017}. The stellar distribution of AGC~114905, consists of an exponential disc with a disc scale length $R_{\rm d} = 1.79 \pm 0.04$~kpc. The left panel of Fig.~\ref{fig:sds} shows its $r-$band stellar image. The galaxy has a relatively blue colour, $(g-r) = 0.3\pm 0.1$ \citep{huds2020,lexi}. We estimate the stellar mass ($M_\ast$) of the galaxy using the mass-to-light-colour relation from \citet{du2020}, which has been recently calibrated using a large sample of low surface brightness galaxies. We obtain $M_\ast = (1.3\pm 0.3)\times 10^8 M_\odot$; this is slightly smaller than the value used in \citet{huds2019,huds2020}, owing to the different mass-to-light-colour calibrations.

We gathered H\,{\sc i} data of AGC~114905 at different resolutions, obtained with the Karl G. Jansky Very Large Array. Specifically, we collected data from the D-, C- and B-array configurations. Details on the D and C configuration observations (PI: Leisman, ID 17A-210) can be found in \citet{leisman2017} and \citet{lexi}. The new B-array observations (PI: Mancera Pi\~na, ID 20A-095) were obtained between July and October 2020. 40 hours were observed (about 34 hours on-source) and combined with the existing C- and D-array observations. The data reduction was done with the software \texttt{Miriad} \citep{miriad} following standard procedures, using a robust weighting of 0.75 to make the final data cube, which results in a cleaned beam of size $7.88$ arcsec $\times$ 6.36~arcsec. After Hanning-smoothing, our final cube has a rms noise per channel of about 0.26~mJy/beam and a spectral resolution of 3.4~km~s$^{-1}$.

The peak H\,{\sc i} column density is 8.4$\times 10^{20}$ atoms~cm$^{-2}$, and the noise level is 4.1$\times 10^{19}$~cm$^{-2}$. The integral flux of the total H\,{\sc i} map is 0.73$\pm 0.07$~Jy~km~s$^{-1}$, close to the value of 0.78~Jy~km~s$^{-1}$ used in \citet{huds2019}, although lower than the $0.96\pm 0.04$~Jy~km~s$^{-1}$ reported by \citet{leisman2017} from unresolved ALFALFA observations. At the distance of AGC~114905, our integral flux yields $M_{\rm HI} = (9.7\pm 1.4)\times 10^8 M_\odot$.\footnote{The ALFALFA flux would instead imply $M_{\rm HI} \approx 1.3\times 10^9 M_\odot$, only strengthening the results shown below.}


We combine $M_{\rm HI}$ and $M_\ast$ to obtain the baryonic mass $M_{\rm bar} = 1.33M_{\rm HI} + M_\ast = (1.4 \pm 0.2) \times 10^9 M_\odot$. The factor 1.33 accounts for the presence of helium, and we neglect any contribution of molecular gas, whose mass is expected to be negligible compared to $M_{\rm HI}$ (e.g. \citealt{hunter2019,CO_udgs}). The galaxy gas fraction $f_{\rm gas} = M_{\rm gas}/M_{\rm bar} \approx 0.9$, ensures that $M_{\rm bar}$ is robust against possible systematics related to $M_\ast$, since the main uncertainty in $M_{\rm gas}$ comes from the distance, which is well constrained.

The total H\,{\sc i} map of the galaxy is shown in the middle panel of Fig.~\ref{fig:sds}, and also on the left panel with the H\,{\sc i} contours overlaid on top of the stellar emission. It is clear that the gas extends well beyond the optical emission, despite UDGs being optically extended. There is also some degree of misalignment between the optical and H\,{\sc i} position angles (see also \citealt{lexi}), although the optical morphology is likely affected by bright, patchy star formation regions. The right panel of Fig.~\ref{fig:sds} shows the surface mass density profiles of the stellar and gaseous discs of AGC~114905. The stellar profile comes from converting our optical surface brightness profiles into mass density using a mass-to-light ratio in the $r-$band of 0.47 \citep{du2020}. The gas profile is obtained using the \texttt{gipsy} \citep{gipsy} task \textsc{ellprof}, and converted to mass density using the conversion factor $1~M_\odot \textrm{pc}^{-2} = 1.25\times10^{20}~\textrm{atoms cm}^{-2}$. Once this conversion is applied, we multiply by the factor 1.33 to account for the presence of helium.

\section{Kinematics}
\label{sec:kinematics}

In order to obtain reliable kinematic information (rotation velocity and velocity dispersion) for our galaxy, we use the software $\mathrm{^{3D}}$Barolo \citep{barolo}. As extensively explained in \citet{barolo}, \citet{enrico_z1}, and \citet{iorio}, $\mathrm{^{3D}}$Barolo\footnote{\url{https://editeodoro.github.io/Bbarolo/}} builds 3D realisations of tilted-ring models of a galaxy data cube, which are then convolved with the beam of the observations and compared channel by channel with the real data. This allows for a robust recovery of the rotation curve and gas velocity dispersion, since the method largely mitigates the effects of beam smearing \citep{bosma1978,begeman,swatersPhD,barolo}.

Before delving into the details and results of our modelling, we will briefly discuss the observed kinematics as well as the derivation of two important geometrical parameters: the position angle of the galaxy and its inclination.

\subsection{Velocity field and geometrical parameters}
\label{sec:inclination}
\begin{figure}
     \centering
     \includegraphics[scale=0.64]{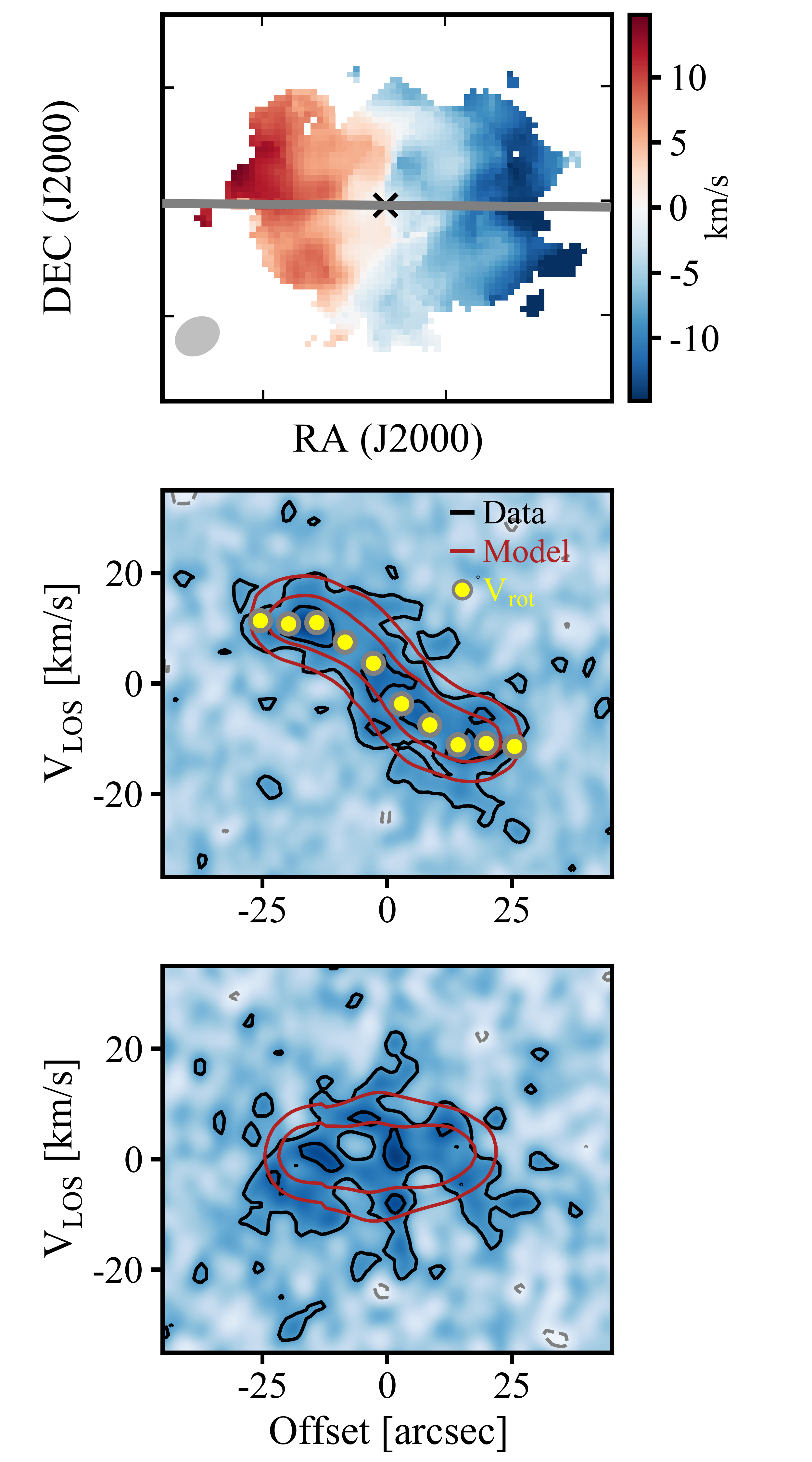}
     \caption{\textit{Top:} Observed velocity field (same physical scale as the total H\,{\sc i} map in Fig.~\ref{fig:sds}); the grey ellipse shows the beam of the observations, the grey line the kinematic major axis, and the black cross the kinematic centre. \textit{Middle (Bottom):} Major (minor)-axis PV diagram; data are shown in blue background and black contours (grey for negative values), and the best-fitting $\mathrm{^{3D}}$Barolo azimuthal model in red contours. The yellow points show the recovered rotation velocities.}
     \label{fig:kinematics}
 \end{figure}

The velocity field (1$^{\rm st}$ moment map) of AGC~114905, shown in the top panel of Fig.~\ref{fig:kinematics}, has the clear kinematic pattern of a regularly rotating disc. The position angle of the velocity field is estimated by trial and error as the angle that maximizes the amplitude of the major-axis position-velocity (PV) diagram (e.g. \citealt{huds2020}). We find a position angle of $89^\circ$, as shown in Fig.~\ref{fig:kinematics} with a line on top of the velocity field. Our value is similar to the $84^\circ$ reported in \citet{huds2019,huds2020} derived from the less resolved data. If we fit the position angle during our kinematic modelling (see below) we find values between $85-92^\circ$ depending on our initial estimates. The middle (bottom) panel of Fig.~\ref{fig:kinematics} shows, in blue background and black contours, the PV diagram along the major (minor) axis of AGC~114905. The major-axis PV shows the typical pattern of a rotating disc and seems to reach a flat velocity in the outer regions. 

While differences of a few degrees in the position angle do not significantly affect the final rotation curve, the inclination of the galaxy is more critical, as small changes at low inclinations can severely affect the value of the deprojected rotation curve. Undoubtedly, the inclination is the main uncertainty in our kinematic modelling and results, and we pay special attention to it. 
Traditionally, the inclination of high-resolution data can be obtained during the kinematic fitting using the velocity field (e.g. \citealt{deblok08}). However, this method is not particularly robust as it depends on the shape of the rotation curve: for solid-body rotation the iso-contours on the velocity field are parallel, nullifying the power to measure the inclination. Given the above, and following \citet{huds2020}, we decide to estimate the inclination with an approach that is independent of the kinematics, relying only on the H\,{\sc i} map of the galaxy.

Our method works as follows. We use $\mathrm{^{3D}}$Barolo to build azimuthal models of the galaxy at different inclinations, with these inclinations being drawn from a flat prior distribution between $10^\circ-80^\circ$ and sampled using a Markov chain Monte Carlo (MCMC) routine (based on the Python package \texttt{emcee}, see \citealt{emcee}). Each model is convolved with the beam of the observations, and its total intensity map is built. We then compare these model intensity maps with the real data, with our MCMC routine minimizing the absolute residuals between model and observed intensity maps. We have tested this method extensively using artificial data cubes matching our resolution and signal-to-noise (S/N), finding it robust and reliable (see also \citealt{filippo_highz}). In the end, we find an inclination of $32 \pm 3^\circ$ for AGC~114905, which we adopt as our fiducial value; the posterior distribution is shown in Fig~\ref{fig:posterior_inc}  in Appendix.~\ref{sec:appendix}.  

As an extra check, we also estimated the inclination in two other ways\footnote{The optical inclination, derived from the optical axis ratio, is around 45$^\circ$. We do not use this inclination in our analysis as it is not clear if the optical data follows the H\,{\sc i} emission (see also e.g. \citealt{lexi,kado-fong_LSBshapes}), but we provide the value for completeness. Clearly, this inclination would lower the value of the circular speed, strengthening our results.}. First, using the method described in \citet{huds2020}, which is equivalent to our method described above but independent of the MCMC sampling. We find an inclination of $34 \pm 5^\circ$, although the quoted uncertainty is just an expected mean value rather than a well defined statistical uncertainty. Similarly, \citet{huds2019,huds2020} found an inclination of $33 \pm 5^\circ$ from their lower-resolution data. Second, we derive kinematic-dependent inclinations. We use both $\mathrm{^{3D}}$Barolo (fitting the whole data cube) and the \texttt{gipsy} task \textsc{rotcur} (fitting the velocity field). Depending on the exact initial value, mask, and ring separation, both methods find inclinations between $30^\circ-37^\circ$. It is reassuring that despite not being our favoured approaches to measure the inclination, we find these different values consistent with the results of our preferred method. Overall, it is important to highlight that we do not find any evidence favouring inclinations lower than 30$^\circ$. 

Finally, it is worth mentioning that while deriving the inclination we assume that the H\,{\sc i} resides on a razor-thin disc, a significant thickness of the disc would imply a higher inclination than what we have derived due to projection effects \citep{iorio_phd}. From this point of view $32 \pm 3^\circ$ gives a lower limit on the inclination of AGC~114905 (see also Sec~\ref{sec:low_inc}). On the other hand, if the gas disc is non-axisymmetric, but instead has some intrinsic elongation, we could be overestimating its inclination. While some simulations suggest this is possible\footnote{\citet{marasco_barsinDM} have reported that about half of the massive dwarf galaxies ($60 < V_{\rm max} < 120$~km~s$^{-1}$, with $V_{\rm max}$ the maximum rotation velocity) in the APOSTLE simulations \citep{apostle_sawala,fattahi_apostle} inhabit dark matter halos with intrinsic axis ratios $b/a < 0.8$; if the disc of AGC~114905 has a similar intrinsic $b/a$ (i.e. it is an elongated disc instead of an inclined circular disc) its inclination could be as low as 10$^\circ$.}, in what follows we assume that the observed HI total intensity map and velocity gradient correspond to an inclined axisymmetric disc galaxy with gas moving in circular orbits.

\subsection{Kinematic modelling}

With the position angle and inclination determined, we proceeded to perform our kinematic modelling with $\mathrm{^{3D}}$Barolo, leaving as free parameters the systemic velocity ($V_{\rm sys}$), the rotation velocity ($V_{\rm rot}$), and the velocity dispersion ($\sigma_{_{\rm HI}}$). We fit an azimuthal model and we use a ring separation of 6~arcsec. This represents a minor oversampling of less than 10 percent with respect to the size of the beam along the major axis of the galaxy (6.5~arcsec), allowing us to trace the rotation curve of the galaxy with five, basically uncorrelated, resolution elements per galaxy side. We also check that the rotation velocities obtained using four or five rings (see below) are well consistent with each other.

We first perform an iteration where $V_{\rm sys}$ is a free parameter. The best $V_{\rm sys}$ turns out to be 5435~km~s$^{-1}$, which matches the centre of a Gaussian fit to the global H\,{\sc i} profile. For our final model we keep this $V_{\rm sys}$ fixed and we only fit $V_{\rm rot}$ and $\sigma_{_{\rm HI}}$. The final model faithfully reproduces the observations. This can be seen in the middle and bottom panels of Fig.~\ref{fig:kinematics}, where we compare the PV diagrams of the best-fitting model (red) and data (black). There are some low--S/N features at low velocities ($\lesssim 5$~km~s$^{-1}$) not reproduced, but $\mathrm{^{3D}}$Barolo closely mimics the overall kinematics of the galaxy. This can also be seen in Appendix~\ref{sec:channelmaps}, where we show representative channel maps of AGC~114905 and of our best-fitting model. The resulting rotation curve, uncorrected for inclination, is shown as yellow points overlaid on top of the major-axis PV diagram of Fig.~\ref{fig:kinematics}, and it is clear that it reaches its flat part well before our outermost radius. To take into account pressure-supported motions, we apply the asymmetric drift correction to our rotation curve (see \citealt{iorio}), ending up with the circular speed $V_{\rm c}$. The correction is found to be very small, contributing less than 2~km~s$^{-1}$ at all radii. 

\begin{figure}
    \centering
    \includegraphics[scale=0.48]{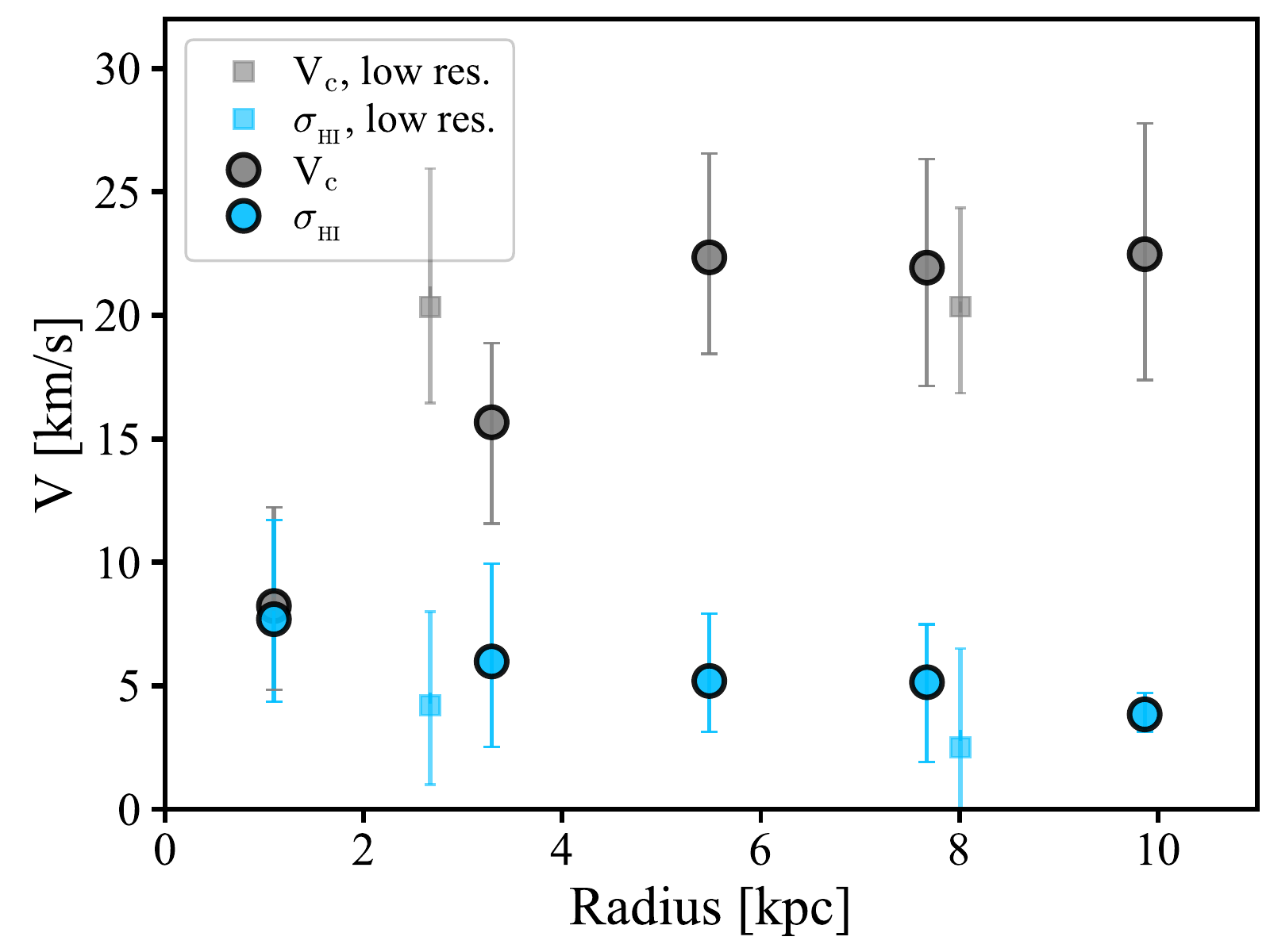}
    \caption{Circular speed (grey circles) and velocity dispersion (blue circles) profiles of AGC~114905, as obtained with our kinematic modelling. Squares show previous results obtained at a lower spatial resolution.}
    \label{fig:rcvdisp}
\end{figure}

In Fig.~\ref{fig:rcvdisp}, we explicitly show the circular speed profile of AGC~114905, as well as its velocity dispersion profile. The uncertainties in $V_{\rm c}$ include the uncertainties in the inclination, by means of a Monte Carlo sampling approach as detailed in \citet{huds2020}. The flat part of the circular speed profile has a velocity of $\simeq 23$~km~s$^{-1}$. This, together with the $M_{\rm bar}$ of the galaxy, confirms its position as an outlier of the BTFR. In Fig.~\ref{fig:rcvdisp} we also include for comparison the values for $V_{\rm c}$ and $\sigma_{_{\rm HI}}$ obtained in \citet{huds2019,huds2020} at lower resolution, showing the good agreement between them and our new determinations. This is important not only for AGC~114905, but for all the UDGs in \citet{huds2019,huds2020}, as it is a direct validation of the lower-resolution results presented previously.

\subsection{Local and global disc stability}

With a median value of $\sim5$~km~s$^{-1}$, the velocity dispersion $\sigma_{_{\rm HI}}$ of AGC~114905 is slightly below the average value in rotation-supported dwarfs ($\sim 8$~km~s$^{-1}$, e.g. \citealt{iorio}) although consistent within the uncertainties. The low values of $V_{\rm c}$ and $\sigma_{_{\rm HI}}$ of our UDG imply a relatively low value of the Toomre parameter $Q_{\rm gas} = \kappa \sigma_{_{\rm HI}}/(\pi G \Sigma_{\rm gas})$, with $\kappa$ the epicycle frequency \citep{toomre1964}. The $Q_{\rm gas}$ profile shows a slight decrease with radius, with a median (mean) value of 0.95 (1.6), after applying a small correction to account for thickness, see \citealt{romeo1994,romeo2013}). The uncertainties are relatively large (typically a factor $2-3$), but these values of $Q_{\rm gas}$ suggest that the galaxy could be susceptible to \emph{local} instabilities (see \citealt{romeo2013} and references therein for a detailed discussion on the interpretation of $Q_{\rm gas}$). While these local instabilities may lead to fragmentation and subsequent star formation, observations suggest this is not always the case \citep{hunter1998,leroy,elmegreen2015}. The value of $Q_{\rm gas}$ for AGC~114905 is lower on average but consistent within $2\sigma$ with the median values of LITTLE THINGS dwarf galaxies \citep{iorio}. Finally, it should be noted that a more detailed calculation that takes into account the gas disc flaring (e.g. \citealt{elmegreen2015,ceci_dwarfs}) would increase the value of $Q_{\rm gas}$, especially in the outer parts.

While $Q_{\rm gas}$ is in principle only related to local instabilities, we can also investigate the \emph{global} disc stability of our UDG. The ordered kinematics seen in Fig.~\ref{fig:kinematics} and the isolation (see \citealt{huds2020}) of the galaxy strongly suggest an equilibrium state. We further tested this by allowing $\mathrm{^{3D}}$Barolo to fit radial motions overlaid on the rotation, but we did not find evidence of such radial motions as their amplitude is always consistent with zero within the uncertainties. We also computed the global stability parameter $X_2 = \kappa^2 R /(4\pi G \Sigma_{\rm gas})$ \citep{toomre1981}, finding a median of 1.2 and with $X_2$ being smaller than 1 (0.9) only at the outermost radius, suggesting the system is stable against bar instabilities ($X_2 \lesssim 1$ is the instability condition often used for dwarf galaxies, see e.g. \citealt{mihos1997,hidalgo-gamez04}). 

Overall, these investigations show that it is reasonable to assume that the cold gas in AGC~114905 is in closed orbits tracing its gravitational potential and allowing us to build mass models based on the derived rotation curve.


\section{Mass modelling}
\label{sec:massmodels}

\subsection{A baryon-dominated rotation curve}

AGC~114905 has a baryonic mass much higher than other dwarf galaxies with similar circular speeds \citep{iorio,huds2020}. It is therefore interesting to see if AGC~114905, like most dwarfs, is dominated by dark matter at all radii.

Prior to obtaining any mass model, we can compare the circular speed profile of the galaxy with the circular speed profile of the baryonic distribution ($V_{\rm bar}$), which is simply the sum in quadrature of the contributions of the stellar and gas discs, this is $V_{\rm bar}^2 = V_{\ast}^2 + V_{\rm gas}^2$. We derive $V_{\rm c,\ast}$ and $V_{\rm c,gas}$ using the software \textsc{galpynamics} \citep{iorio_phd}. \textsc{galpynamics}\footnote{\url{https://gitlab.com/iogiul/galpynamics/}} takes as input the mass density profile of a given component, fitted with an appropriate function (see below), computes its gravitational potential via numerical integration, and returns the associated circular speed. 

In the case of the stellar disc, we use an exponential profile with $M_\ast = 1.3\times10^8~M_\odot$ and an exponential disc scale length $R_{\rm d} = 1.79$~kpc; this profile can be compared with the data in Fig.~\ref{fig:sds}. We assume a sech$^2$ profile along the vertical direction, and a constant thickness $z_{\rm d} = 0.196 R_{\rm d}^{0.633} \approx$~280~pc, as found in low-inclination star forming galaxies \citep{bershady_thickness}.

For the gas component (H\,{\sc i} plus helium), we fit the density profile with a profile of the form
\begin{equation}
    \Sigma_{\rm gas} = \Sigma_{\rm 0, gas} e^{-R/R_1} (1+r/R_2)^\alpha~,
\end{equation}
where $\Sigma_{\rm 0,gas}$ is the gas central surface density, $R$ is the cylindrical radius, and $R_1, R_2$, and $\alpha$ are the fitting parameters (equal to $3.2~M_\odot/\textrm{pc}^2$, $1.1$~kpc, $16.5$~kpc, and $18$, respectively). This profile provides a good fit to the observations, as seen in Fig.~\ref{fig:sds}. For the vertical structure of the gaseous disc we assume a Gaussian profile and a constant vertical scale-height $z_{\rm d} = 250$~pc. It is worth mentioning that the results we show below do not depend significantly on the assumed thickness of the stellar or gaseous discs.

\begin{figure}
    \centering
    \includegraphics[scale=0.5]{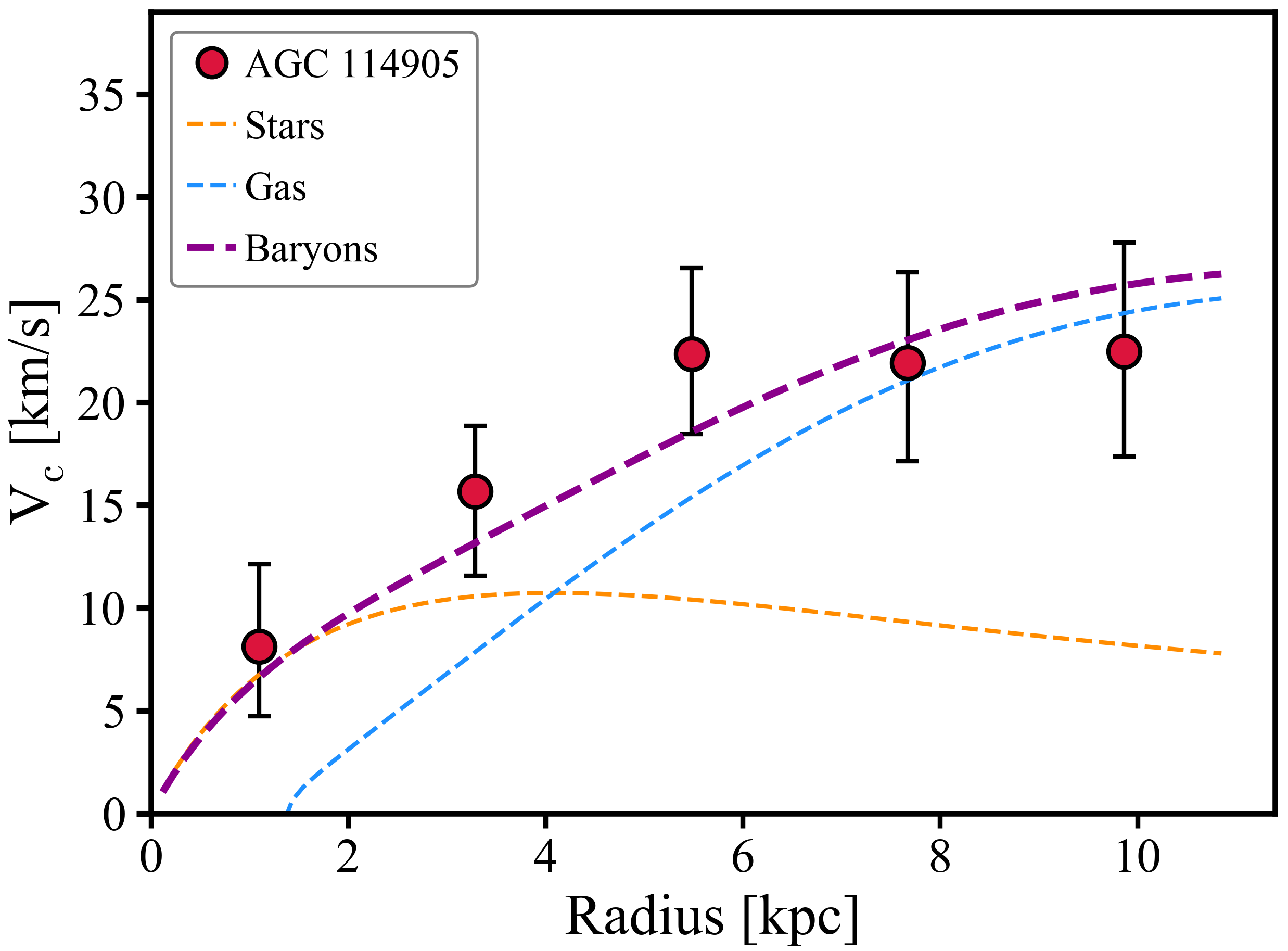}
    \caption{Circular speed profile of AGC~114905 (red points) compared to the contribution expected from stars (orange line), gas (blue line), and baryons (stars plus gas, magenta line).}
    \label{fig:baryondominated}
\end{figure}

Fig.~\ref{fig:baryondominated} shows the contribution of $V_{\ast}$, $V_{\rm gas}$, and $V_{\rm bar}$ to the total $V_{\rm c}$ of AGC~114905. Remarkably, $V_{\rm bar}$ provides a reasonable description of $V_{\rm c}$ at all radii. This implies that as opposed to classical dwarf galaxies (e.g. \citealt{iorio,read2017}), the dynamics of AGC~114905, at least within the observed radii extending to about 10~kpc, are baryon-dominated rather than dark-matter dominated. This was already postulated in \citet{huds2019}, but it is now confirmed with a well traced rotation curve.

\subsection{Fitting cold dark matter halos}

In our current framework of galaxy formation, we expect every galaxy to be embedded in a cold dark matter (CDM) halo. Because of this, it is relevant to investigate whether or not physically motivated CDM haloes can be consistent with our rotation curve, even if Fig.~\ref{fig:baryondominated} suggests the absence of a dynamically significant halo in AGC~114905. We aim to find a dark matter halo whose circular speed $V_{\rm DM}$ meets
\begin{equation}
\label{eq:massmodel}
    V_{\rm c}^2 = V_\ast^2 + V_{\rm gas}^2 + V_{\rm DM}^2~.
\end{equation}

Dark matter haloes are often described with the so-called NFW profile \citep{nfw}, whose density as a function of the spherical radius in cylindrical coordinates $r$ ($r~=~\sqrt{R^2+z^2}$~) is given by
\begin{equation}
    \rho_{\rm NFW}(r) = \dfrac{4 \rho_{\rm s}}{(r/r_{\rm s})(1 + r/r_{\rm s})^2}~,
\end{equation}
where $r_{\rm s}$ is a `scale radius' and $\rho_{\rm s}$ is the density at $r_{\rm s}$. We will denote the corresponding mass profile as $M_{\rm NFW}(r)$:
\begin{equation}
    M_{\rm NFW}(<r) = \dfrac{M_{200}}{\ln(1+c_{200}) - \dfrac{c_{200}}{1+c_{200}}}\left[\ln\left( 1 + \dfrac{r}{r_{\rm s}}\right) - \dfrac{r}{r_{\rm s}} \left( 1 + \dfrac{r}{r_{\rm s}}\right)^{-1} \right]~.
\end{equation}
The parameter $M_{200}$ is defined as the mass within a sphere with radius $r_{200}$ within which the average density is 200 times the critical density of the universe, while the concentration $c_{200}$ is defined as $c_{200} = r_{200}/r_{\rm s}$.

While NFW haloes provide good descriptions for massive galaxies, this is not the case for dwarf galaxies (see \citealt{bullock2017}). Therefore, for our UDG, we assume that the dark matter halo is described by a `\textsc{core}NFW' profile \citep{coreNFW}, which is an extension of the classical NFW profile that has the flexibility to develop -or not- a core. In Section~ \ref{sec:low_c200} we discuss other halo profiles. The \textsc{core}NFW profile has been found to fit very well rotation curves of dwarf galaxies, both real and simulated \citep{coreNFW,readAD}.

The density profile of the \textsc{core}NFW halo can be written as
\begin{equation}
    \rho_{\rm coreNFW}(r) = f^n \rho_{\rm NFW}(r) + \dfrac{nf^{n-1}(1-f^2)}{4\pi r^2 r_{\rm c}} M_{\rm NFW}(r)~.
\end{equation}
Here, $\rho_{\rm NFW}$ and $M_{\rm NFW}$ are the above NFW parameters, while $f$ is a function (defined as $f~=~\tanh(r/r_{\rm c})$) that generates a core of size $r_{\rm c}$. In principle, $r_{\rm c}$ can be a fitting parameter, but as discussed in detail by \citet{coreNFW,readAD,read2017}, fixing it to $r_{\rm c} = 2.94~R_{\rm d}$\footnote{In principle, $r_{\rm c} = 1.75~R_{\rm e}$, with $R_{\rm e}$ the half-light radius. For an exponential profile ($R_{\rm e} = 1.678 R_{\rm d}$) this becomes $r_{\rm c} = 2.94~R_{\rm d}$.} is in good agreement with simulations and observations where $r_{\rm c}$ is fitted as free parameter. Importantly, the factor 2.94 cannot be significantly larger as there is not enough energy from supernovae to create cores of size much larger than $2.94~R_{\rm d}$ (see also e.g. \citealt{benitezllambay19,coreEinasto,trujillogomez_udgs}). The degree of transformation from cusp to core is described by the parameter $n$, with $n=0$ defining a cuspy NFW profile and $n = 1$ a completely cored profile. The parameter $n$ is defined as $n = \tanh(\kappa t_{\rm SF} / t_{\rm dyn})$, with $\kappa = 0.04$ a fixed parameter, $t_{\rm SF}$ the time whilst the galaxy has been forming stars (set to 14~Gyr), and $t_{\rm dyn}$ the NFW dynamical time at the scale radius $r_{\rm s}$, which can be expressed in terms of $M_{\rm NFW}$, $r_{\rm s}$, and $G$ (the Newtonian gravitational constant) as 
\begin{equation}
\label{eq:tdyn}
    t_{\rm dyn} = 2 \pi \sqrt{\dfrac{r_{\rm s}^3}{G M_{\rm NFW}(<r_{\rm s})}} ~~~.
\end{equation}

The dark matter profile has then the same two free parameters as a NFW halo: the mass of the halo ($M_{200}$) and its concentration ($c_{200}$). N-body cosmological simulations find a strong correlation between $c_{200}$ and $M_{200}$ (e.g. \citealt{duttonmaccio2014,ludlow2014}), so in practice one can even fit NFW-like profiles with one single parameter. While the the other parameters of the halo ($n$, $r_{\rm s}$, $r_{\rm c}$), are not considered free parameters, they also change on each step of the MCMC, as they depend on $M_{200}$ and $c_{200}$ as described above. 

To find the best-fitting CDM halo we use a MCMC routine (also based on \texttt{emcee}) that minimises the residuals of Eq.~\ref{eq:massmodel} using a standard $\exp(-0.5 \chi^2)$ function as likelihood, with $\chi^2$ given by
\begin{equation}
    \chi^2 = \Large{\sum} \dfrac{(V_{\rm c}-V_{\rm c,mod})^2}{\delta_{V_{\rm c}}^2}~,
\end{equation}
where $V_{\rm c}$ and $V_{\rm c,mod}$ are the observed and model circular speed profiles, respectively, and $\delta_{V_{\rm c}}$ is the $\mathrm{^{3D}}$Barolo uncertainty in the kinematic modelling, which we assume to be Gaussian (see \citealt{enrico_phd}). As we discuss next, the inclination is a free parameter in our MCMC, and thus $\delta_{V_{\rm c}}$ itself does not include the contribution from the inclination uncertainty; these error bars are, therefore, smaller than those shown in Figs.~\ref{fig:rcvdisp} and \ref{fig:baryondominated}.

The MCMC explores the ($M_{200}, c_{200}$) parameter space and retrieves the best-fitting combination. In addition to $M_{200}$ and $c_{200}$, we include the distance $D$ and the inclination $i$ as nuisance parameters. In practice, we impose a Gaussian prior on $D$ centered at 76~Mpc and with a standard deviation of 5~Mpc, exploring the $2\sigma$ range $66 \leq D/\textrm{Mpc} \leq 86$. Similarly, for $i$, we impose a Gaussian prior centered at $32^\circ$ with a standard deviation of $3^\circ$, within $26^\circ \leq i \leq 38^\circ$; in Section~ \ref{sec:discussion} we also discuss the case where the priors for $D$ and $i$ are wider. It is worth pointing out that a change in $D$ introduces a change in the conversion factor between arcsecond and kpc, thus modifying our sampling of the rotation curve. Additionally, it affects the value of $R_{\rm d}$, which in turn changes $r_{\rm c}$ and the thickness of the stellar disc. On the other hand, $i$ affects the overall normalization of the rotation curve and of the gas circular speed profile. Having established this, we explore different scenarios, which differ by our chosen priors on $M_{200}$ and $c_{200}$. 

In a very first attempt, we use the flat priors $6 \leq \log(M_{200}/M_\odot) \leq 12$ and $0.1 \leq c_{200} \leq 30$. However, $c_{200}$ remains completely unconstrained as its posterior distribution is flat over all the explored range. Upon imposing the $c_{200}-M_{200}$ relation of \citet{duttonmaccio2014} as a Gaussian prior on $c_{200}$, we find $\log(M_{200}/M_\odot) = 7.6^{+0.7}_{-1.0}$, $c_{200} = 21^{+5}_{-3}$, $D = (76 \pm 5)$~Mpc, and $i = 33\pm3^{\circ}$. 

While the resulting fit is close to the data (given that $V_{\rm bar} \approx V_{\rm c}$ and $V_{\rm DM}$ is subdominant), the value of $M_{200} \sim 10^8~M_\odot$ is too low to be plausible in a CDM cosmology. Given $M_{\rm bar} = 1.4\times 10^9~M_\odot$, the very minimum expected $M_{200}$ (assuming the galaxy has a baryon fraction as high as the cosmological average: $f_{\rm c,bar} \simeq 0.16$, see \citealt{bookFilippo}) is about $0.9\times 10^{10} M_\odot$. It is clear that the MCMC routine finds a low mass since $V_{\rm c} \simeq V_{\rm bar}$, but the resulting halo does not seem to have a physical justification.

Taking the above into consideration, we decided to impose a lower boundary to the prior of $M_{200}$ such that the minimum halo would produce $M_{\rm bar}/M_{200} \simeq 0.16$ (i.e. the cosmological baryon fraction). With this, the prior for $M_{200}$ becomes $10 \leq \log(M_{200}/M_\odot) \leq 12$. We stress that the lower limit on the prior corresponds to the minimum expected value of $M_{200}$. In theory one expects the galaxy to have a significantly larger $M_{200}$. For example, the $\Lambda$CDM stellar-to-halo mass relation from \citet{posti_galaxyhalo} would predict $\log(M_{200}/M_\odot) \approx 10.6$. For $c_{200}$ we explore two scenarios: one where we impose again the $c_{200}-M_{200}$ relation of \citet{duttonmaccio2014} as a Gaussian prior in the MCMC routine, and one where $c_{200}$ has a flat wide prior $0.1 \leq c_{200} \leq 30$. In what follows, we refer to these two scenarios as Case 1 and Case 2, respectively.

The posterior distributions of both Case 1 and Case 2 are shown in Appendix~\ref{sec:appendix}. Somewhat unsurprisingly, for both cases the MCMC finds $\log(M_{200}/M_\odot) \simeq 10$, with posterior distributions that simply try to go to the lower bound. In Case 1, (imposing the Gaussian prior on the $c_{200}-M_{200}$ relation), we find $D = (67\pm 1)$~Mpc and $i = 26.1^{\circ +0.2}_{~-0.1}$, with posterior distributions also trying to go towards their lower bounds (see Fig.~\ref{fig:posterior_case1}). The concentration, on the other hand, is well constrained, and we find $c_{200} = 11.7 \pm 0.3$. The other parameters of the \textsc{core}NFW profile are $n = 0.7$, $r_{\rm c} = 4.6$~kpc, and $r_{\rm s} = 3.8$~kpc. 

For Case 2, while $D$ and $i$ are well constrained following their priors ($D = 73\pm4$~Mpc, $i = 29^\circ \pm 2^\circ$), the posterior distribution of $c_{200}$ goes to its lower bound, $c_{200} = 0.3^{+0.3}_{-0.2}$ (see Fig.~\ref{fig:posterior_case2}). The other parameters of the \textsc{core}NFW profile are $\log(M_{200}/M_\odot)=10.2$, $n = 0.03$ (i.e. \textsc{core}NFW $\approx$ NFW) driven by a high $t_{\rm dyn}$ (Eq.~\ref{eq:tdyn}), $r_{\rm c} = 5.1$~kpc, and $r_{\rm s} = 166$~kpc, driven by the extremely low $c_{200}$ ($r_{\rm s} = r_{200}/c_{200}$).

\begin{figure*}
    \centering
    \includegraphics[scale=0.49]{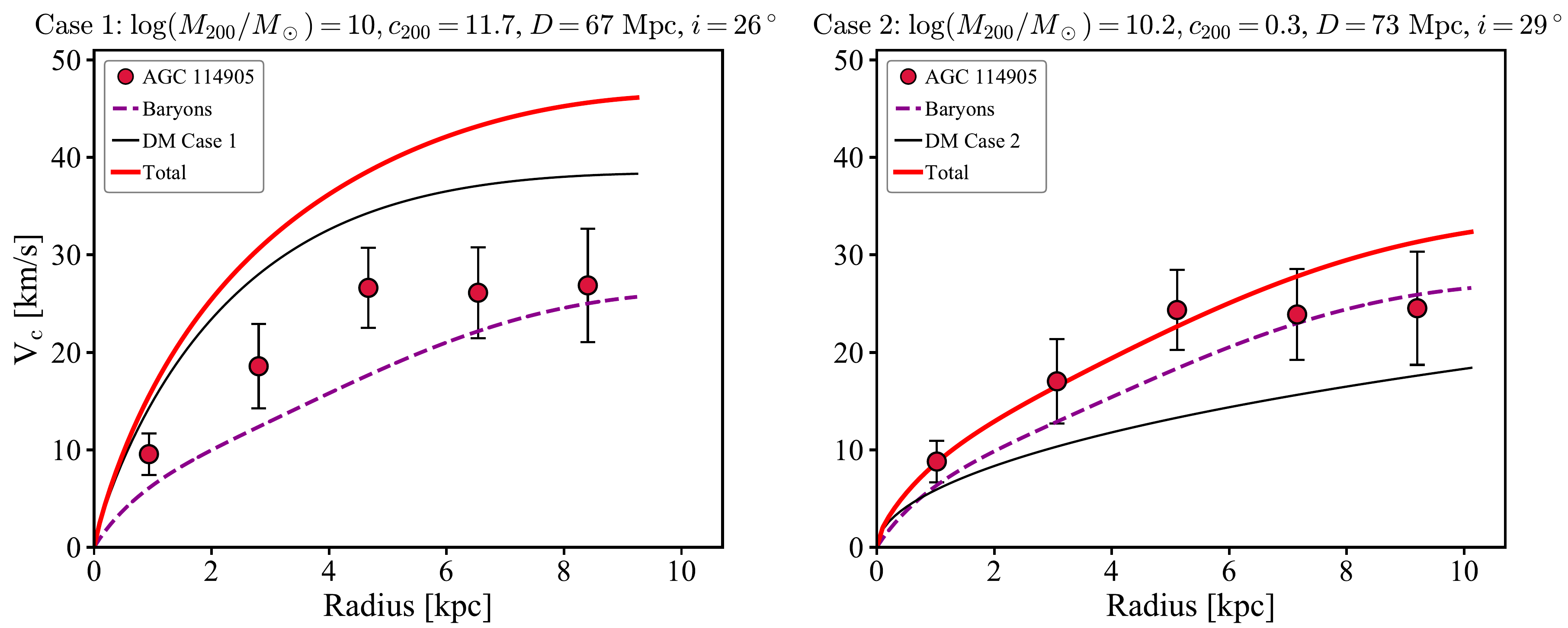}
    \caption{Mass models of AGC~114905. Case 1 and Case 2 are shown on the left and right panels, respectively. In both panels the red points show the $V_{\rm c}$ profile of AGC~114905, while the dashed magenta lines represent the $V_{\rm c}$ expected from the baryons (stars plus gas). The dark matter haloes are shown with black lines, and the red lines give the total contribution of baryons and dark matter together. Case 1, which follows the CDM $c_{200}-M_{200}$ relation, is inconsistent with the observations. Case 2 fits the data better, but it has a $c_{200}$ too low for CDM. Note also that the assumed distance and inclination are different between both panels. Because the assumed distance is different on each panel, the sampling of the rotation curve along the horizontal axes is also different. In a similar way, the normalization of the rotation curves differ from each other due to the different inclinations. See the text for details.}
    \label{fig:massmodels}
\end{figure*}

Fig.~\ref{fig:massmodels} shows the two resulting mass models. Case 1, on the left panel, is in clear disagreement with the data as it significantly overestimates $V_{\rm c}$, even when the distance and inclination go to their lowest allowed values. Case 2, on the right panel, lies closer to the data but presents other problems, as we discuss in the next Section. 


\section{Discussion}
\label{sec:discussion}

Having presented our main results, we now discuss their implications. Provided our rotation curve for AGC~114905 faithfully traces its circular speed, the fact that it is baryon-dominated out to the outermost observed radius (Fig.~\ref{fig:baryondominated}) implies two possible scenarios is a $\Lambda$CDM context: that our UDG lacks a significant amount of dark matter across all radii (even beyond the range probed by our data), or that it has a peculiar dark matter halo with little mass within 10~kpc (e.g. right panel in Fig.~\ref{fig:massmodels}).

\subsection{AGC~114905 compared to `dark-matter free' galaxies}

Since \citet{vandokkum_DF2} and \citet{vandokkum_DF4} postulated that DF--2 and DF--4 have very low or no dark matter content, different mechanisms to create such peculiar galaxies have been proposed. One of the leading ideas is that high-velocity ($\sim 300~\textrm{km s}^{-1}$) collisions between gas-rich dwarf galaxies can create dark-matter free (or almost dark-matter free, $M_{\rm DM} \sim 10^5 M_\odot$) galaxies \citep{silk_udgs,shin_udgs}. Importantly, those types of galaxies are expected to form in dense environments and to have a baryonic mass dominated by stars rather than cold gas. Another mechanism proposed to explain the existence of DF--2 and DF--4 are tidal interactions with massive neighbouring galaxies that strip the dark matter away \citep{jackson_tidalUDGs,doppel_udgs}; \citet{montes_df4} claim that in fact DF--4 currently shows signs of such interactions. 

While the above scenarios can manage to produce dark-matter poor, UDG-like galaxies that show some degree of similarity with DF--2 and DF--4, it is important to bear in mind that gas-rich UDGs are rather different objects. Not only are they gas-dominated ($f_{\rm gas} \simeq 0.9$ for AGC~114905), but they are also isolated (by selection, see \citealt{leisman2017}). In the specific case of AGC~114905, the nearest galaxy within a recession velocity of 500~km~s$^{-1}$ with confirmed (optical or HI) redshift, is the faint dwarf AGC~114806 at a projected distance of 2.1~Mpc and with a systemic velocity within a few km~s$^{-1}$. Using data from the Sloan Digital Sky Survey \citep{sdss_dr12}, we also looked for possible unconfirmed massive companions of AGC~114905. We explored the area within 45~arcmin of AGC~114905, corresponding to a circle of radius 1~Mpc at the distance of AGC~114905, querying for galaxies with $R_{\rm e} \geq 1~\textrm{kpc}$ and with color $(g-r) 
\leq 1~\textrm{mag}$. In this region, there are only seven galaxies with $M_\ast \gtrsim 10^9 M_\odot$ (assuming a distance of 76~Mpc and the mass-to-light-colour relation from \citealt{du2020}) with unknown distance. All of them resemble confirmed background red galaxies, and the closest in projection lies at 700~kpc. All this evidence, together with the lack of tidal features in the optical and H\,{\sc i} morphology of our UDG, suggests that it is truly isolated.

An idea that could reconcile a tidal origin with the current isolation of AGC~114905 is that it is an old tidal dwarf galaxy (TDG, e.g., \citealt{duc_tdgs}), since TDGs are expected to have a low dark matter content and low rotation velocities (e.g. \citealt{hunter_tdgs2000,lelli_tdgs}). If the interaction that originated the TDG happened at high redshift ($z \sim 4-6$) and the galaxy had an escape velocity of $\sim 400$~km~s$^{-1}$, the parent galaxy would lie today at distances about $\sim 5$ Mpc from AGC~114905. While this scenario is impossible to test in practice, the population of known old TDGs in the nearby universe both in observations and simulations are found at much closer distances and recessional velocities from their parent galaxies than what AGC~114905 (and the similar gas-rich UDGs from \citealt{huds2019,huds2020}) is from any massive galaxy \citep{hunter_tdgs2000,kaviraj_tdgs,duc_tdgs,haslbauer_UDGsnoDMIllustris}. Overall, while is difficult to give a final answer, it seems unlikely that the small (if any) amount of dark matter in AGC~114905 can be attributed to the above mentioned mechanisms perhaps valid for DF--2 and DF--4. 

Recently, \citet{trujillogomez_udgs} proposed a semi-empirical model where strong feedback from globular clusters can produce UDGs with dark matter cores as large as $10-30$~kpc. However, the model does not include a detailed treatment of the gas component which is the dominant mass budget of gas-rich UDGs, and a thorough comparison with our data is not yet possible to carry out.

It would be instructive to obtain information about the kinematics of AGC~114905 beyond 10~kpc, where the contribution of stars and gas becomes smaller and would produce a declining rotation curve. Instead, if a flat rotation curve were to be found, it would suggest the presence of dark matter. In the next Section, motivated by our results in Fig.~\ref{fig:massmodels}, we discuss which type of CDM haloes are in agreement or disagreement with our observations.

\subsection{The $c_{200}$ of a CDM halo for AGC~114905 is too low}
\label{sec:low_c200}

It follows from Fig.~\ref{fig:massmodels} that it does not seem possible to fit the circular speed profile of AGC~114905 with a CDM-motivated $c_{200}$. As mentioned above, if the $c_{200}-M_{200}$ relation is imposed (Case 1, left panel of Fig.~\ref{fig:massmodels}), it fails by a large margin at reproducing the amplitude of the circular speed profile. Case 2 (right panel of Fig.~\ref{fig:massmodels}), fitting a free $c_{200}$, does a better job and it is consistent with the circular speed profile within the uncertainties. However, $c_{200}$ is too low and completely off the expected $c_{200}-M_{200}$ relation that emerges from cosmological simulations\footnote{Assuming that the scatter of the $c_{200}-M_{200}$ relation measured at high masses ($\sigma_{\log(c_{200})}~=~0.11$~dex, see \citealt{duttonmaccio2014})  is applicable also at $M_{200} \lesssim 10^{10}~M_\odot$, then $c_{200}$ of AGC~114905 is about 15$\sigma$ below the expected value, although this number could be reduced if the $c_{200}-M_{200}$ or its scatter depart from Gaussian (Kong, D. et al. in prep.).} (e.g. \citealt{duttonmaccio2014,ludlow2014}) and it might even be non-physical: \citet{mcgaugh2003} argue that CDM haloes with $c_{200} < 2$ are not produced in any sensible cosmology. \citet{sengupta2019} and \citet{shi2021} also suggested that the gas-rich UDGs AGC~242019 and UGC~2165, respectively, have a $c_{200}$ around 2. Drawing conclusions from those galaxies might be less straightforward: the rotation curve of AGC~242019 does not seem to reach its flat part, while the data of UGC~2165 have low resolution and its rotation curve (apparently rising as solid-body) is significantly oversampled. Still, it is interesting that similarly low values of $c_{200}$ are reported. 

It is important to highlight that while low surface brightness galaxies have been historically found to inhabit low-concentration haloes (e.g. \citealt{mcgaugh2003}), the concentrations of those haloes are still usually in broad to good agreement with $\Lambda$CDM cosmology \citep{maccio2007}, while the concentration of AGC~114905 is rejected at a high significance level. Given the volume explored by \citet{leisman2017} when building the parent sample of AGC~114905 ($\sim 10^6$~Mpc$^3$, see \citealt{alfalfa,jones_huds}), finding a single galaxy with the properties of AGC~114905 should be practically impossible in a CDM Universe. This result becomes even stronger considering the rest of the sample studied in \citet{huds2019,huds2020} possibly shows similar properties, even if slightly less extreme as AGC~114905 presents the largest offset from the BTFR. In this context, it is important to bear in mind that gas-rich UDGs as a whole population have significantly narrower velocity widths (a proxy for their rotation velocities) than galaxies of similar mass \citep{leisman2017,jones_huds}.

We also note here that the implausibility of the CDM halo needed in AGC~114905 is not just related with the cusp-core problem \citep{bullock2017}, since by fitting a \textsc{core}NFW profile we do not force the halo to be cored or cuspy per se. It is also clear that the scales at which dark matter is deficient in AGC~114905 (10~kpc) are larger than any realistic core size for dwarf galaxies in both observations and simulations (e.g. \citealt{coreNFW,read2017,coreEinasto}). To further explore this, we performed a run of our MCMC routine where $r_{\rm c}$ is kept as a free parameter. In practice, we use a flat prior exploring the range $0 \leq r_{\rm c}/\textrm{kpc} \leq 44$. The maximum value of 44~kpc is chosen because it is the value of $r_{200}$ given $M_{200} = 10^{10}~M_\odot$. Additionally, we impose a minimum value on the prior of $M_{200}$, $\log(M_{200}/M_\odot)=10$, as well as the Gaussian prior on the $c_{200}-M_{200}$ relation. The MCMC routine finds the parameters $\log(M_{200}/M_\odot) \approx 10^{10}$, $c_{200} \approx 12$, $D \approx 71$~Mpc, $i \approx 27^\circ$, and $r_{\rm c} \approx 41$~kpc, with the $i$ and $r_{\rm c}$ posterior distributions simply going to their minimum and maximum allowed values, respectively. While the fit is just slightly worse than Case 2 (right panel of Fig.~\ref{fig:massmodels}) it seems non-physical. Expressing the core radius as $r_{\rm c} = \eta R_{\rm e}$ implies $\eta \sim 15$. As discussed by \citet{read2017}, there is not enough supernovae energy in galaxies to drive  $\eta > 2.75$, and $\eta = 1.75$ fits real and simulated galaxies well. Even if other energy sources (e.g. \citealt{bookFilippo}) can affect the distribution of dark matter in galaxies, it seems unlikely that they would contribute much more than supernovae, as required to achieve $r_{\rm c} \approx r_{200}$.

Finally, it is worth clarifying that the problem of fitting a CDM halo to AGC~114905 is not restricted to specific functional forms such as the \textsc{core}NFW profile. In addition to \textsc{core}NFW, we also try with the \textsc{core}Einasto halo, which allows the Einasto profile to develop a core, and which has been found by \citet{coreEinasto} to successfully reproduce the cored dark matter profile of a variety of galaxies in the FIRE-2 simulations \citep{fire2}. For this profile we also impose a minimum $\log(M_{200}/M_\odot)=10$, but the only way to find agreement with our data is again if the size of the core is as large as $r_{200}$. 

Overall, the existence of galaxies like AGC~114905 seems to pose a major challenge for CDM haloes. An interesting line of research is to explore whether or not the current issues can be mitigated by invoking a different type of dark matter (e.g. \citealt{kaplinghat2020,yang_sidm}).


\subsection{AGC~114905 challenging MOND}
\label{sec:mond}
Modified Newtonian dynamics (MOND, \citealt{mond,sanders_mond_review}) is an alternative approach to dark matter theories which aims to explain dark matter physics by invoking a modification to Newtonian dynamics. One of the major achievements of MOND is how well it predicts rotation curves of galaxies (e.g. \citealt{sanders_mond_review,mond_famaey_review} and references therein). Rather than a fit, MOND makes a direct prediction of the shape of the rotation curve given the stellar and gas mass distributions. The only other parameter is $a_0 \simeq 1.2\times 10^{-8}$~cm~s$^{-2}$, postulated to be a universal constant.

The fact that some isolated UDGs are off the BTFR already posed a challenge to MOND, which predicts a tight (zero intrinsic scatter) BTFR with slope of 4 for isolated galaxies. Yet, some doubts about this may exist, as it could be argued that the circular speeds reported in \citet{huds2019,huds2020} were not tracing the flat part of the rotation curve or were too affected by the resolution; here we have shown that this is not the case for AGC~114905. 

Following \citet{gentile2008}, in the absence of an `external' gravitational field of a neighbouring massive galaxy, the MOND rotation curve can be written as 
\begin{equation}
  V^2_{\rm MOND}(r) =  V_{\rm bar}^2  + \dfrac{V_{\rm bar}^2}{2} \left(\sqrt{1+\dfrac{4 a_0 r}{V_{\rm bar}^2}}-1 \right)~.
\end{equation}

\begin{figure}
    \centering
    \includegraphics[scale=0.5]{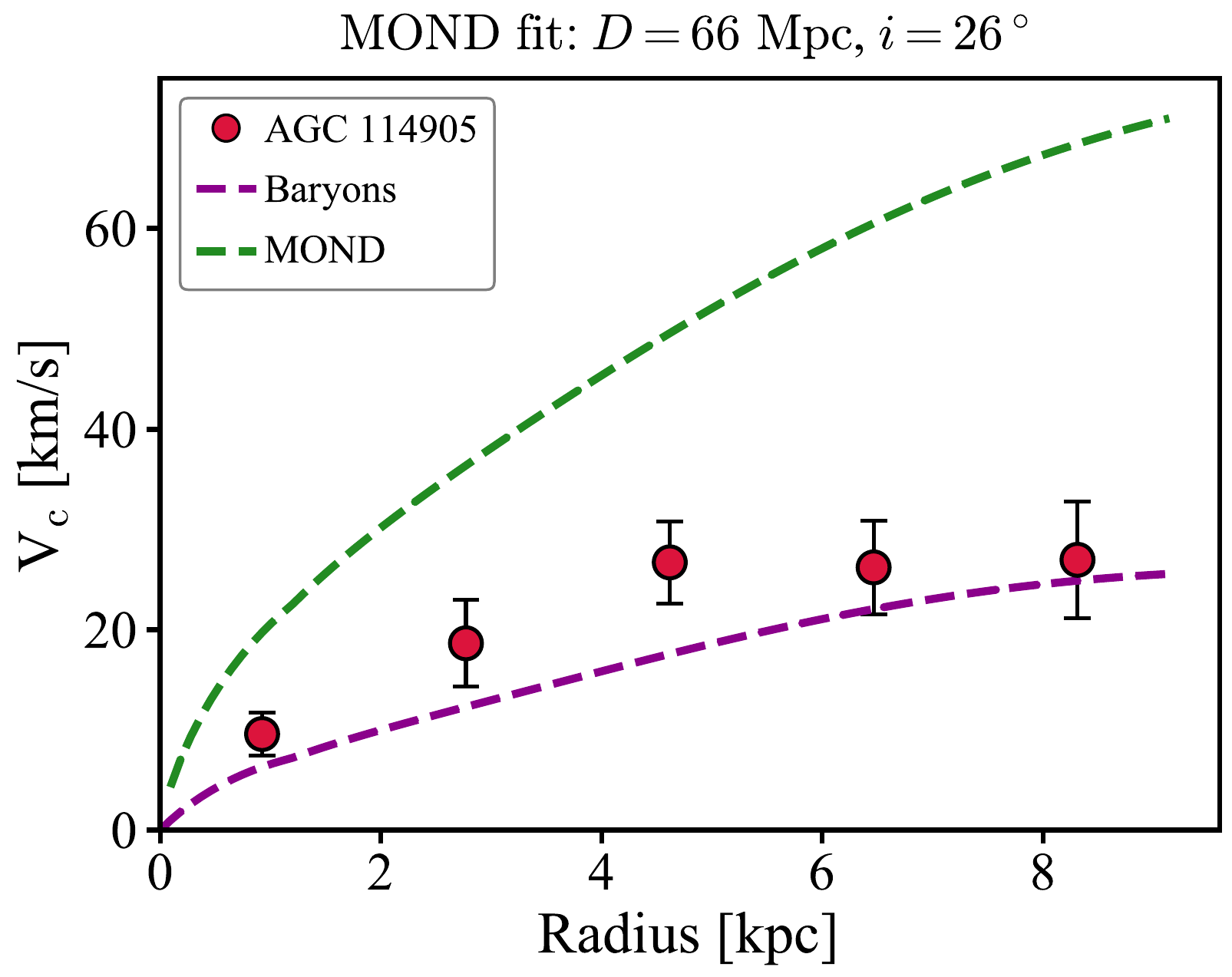}
    \caption{MOND prediction (green line) of the circular speed profile of AGC~114905 (red points). The baryonic circular speed profile is shown with a magenta line.}
    \label{fig:MOND}
\end{figure}

We stress that the above formula is applicable to our UDG given the lack of massive galaxies in its vicinity. In order to test this prediction, we performed another MCMC fit with $D$ and $i$ as free parameters (which affect both $V_{\rm c}$ and $V_{\rm MOND}$), following the same Gaussian priors as for our Case 1 and Case 2 above. Even with the posterior distributions for $D$ and $i$ going to their lower bounds ($D \approx 66$~Mpc and $i \approx 26^\circ$, see Fig.~\ref{fig:posterior_mond} in Appendix~\ref{sec:appendix}), the MOND prediction markedly overestimates the circular speed of our UDG (consistent with the offset from the BTFR), as we show in Fig.~\ref{fig:MOND}. There may be also some tension with the shape of the rotation curve, which is not predicted to be flat as in the observations. Therefore, our UDG seemingly presents a challenge to MOND, which can only be reconciled by invoking a much lower inclination, as we discuss in Section~ \ref{sec:low_inc}. 

\subsection{The effects of a lower inclination}
\label{sec:low_inc}
As described in Section~ \ref{sec:kinematics}, we measure the inclination of AGC~114905 to be $32^\circ \pm 3^\circ$, using a well tested method that relies exclusively on the total H\,{\sc i} map and is independent of the kinematics of the galaxy and our posterior kinematic modelling. While we argue that our inclination is robust, our results are certainly dependent on it. In particular, if AGC~114905 had a much lower inclination, the amplitude of its rotation curve would be significantly larger, having more room for dark matter within the observed radii and potentially alleviating some of the tensions presented in this paper (see for instance the case of IC~1613 in \citealt{oman_missingdarkmatter}).

Given this, it is interesting to quantify by how much the inclination of our UDG would need to decrease to make it consistent with the CDM (and MOND) expectation. For this exploration we assume a \textsc{core}NFW profile with $r_{\rm c} = 2.94~R_{\rm d}$. The `minimum' expectation is such that the galaxy has $\log(M_{200}/M_\odot) = 10$ and a $c_{200}$ in agreement (within some scatter) with the $c_{200}-M_{200}$ relation. We run again our MCMC routine, but this time we use wider Gaussian priors for $D$ and $i$: $50 \leq D/\textrm{Mpc} \leq 100$ and $5^\circ \leq i < 85^\circ$, with centre and standard deviations as in Case 1. The resulting parameters are $\log(M_{200}/M_\odot) = 10.03^{+0.06}_{-0.03} M_\odot$, $c_{200} = 12.1 \pm 0.3$, $D = (72.7 \pm 5)$~Mpc, and $i = (15\pm 1)^\circ$. The low inclination brings up the circular speed of AGC~114905 to velocities around 45~km~s$^{-1}, $which are consistent with a \textsc{core}NFW halo similar to the halo in Case 1 (the other parameters are $n=0.73$, $r_{\rm c} = 5$~kpc, and $r_{\rm s} = 3.8$~kpc).

With an inclination of $15^\circ$ the galaxy would be still quite puzzling, as it would have a baryon fraction about 70 per cent the value of the cosmological average (90 per cent if $D = 76$~Mpc), as opposed to most dwarf galaxies that have low baryon fractions of a few per cent \citep{mcgaugh_baryonfraction,read2017}. Additionally, AGC 114905 would still lie off the BTFR. If one instead imposes $\log(M_{200}/M_\odot) = 10.6$ (which gives a baryon fraction of about 20 per cent) assuming the $\Lambda$CDM stellar-to-halo mass relation from \citet{posti_galaxyhalo}, a corresponding $c_{200} = 11.5$ (following the $c_{200}-M_{200}$ relation), and $D = 76$~Mpc, the needed inclination is $10.4 \pm 0.3^\circ$. Similarly, an inclination of $i = 10.8 \pm 0.3^\circ$ would be needed in order to find agreement between the MOND prediction and the $V_{\rm c}$ profile of AGC~114905, at least on average, since the shape predicted by MOND seems to also differ from our rotation curve. Note, however, that a radially varying inclination could potentially alleviate this tension between the rotation curves shapes. 

\begin{figure}
    \centering
    \includegraphics[scale=0.47]{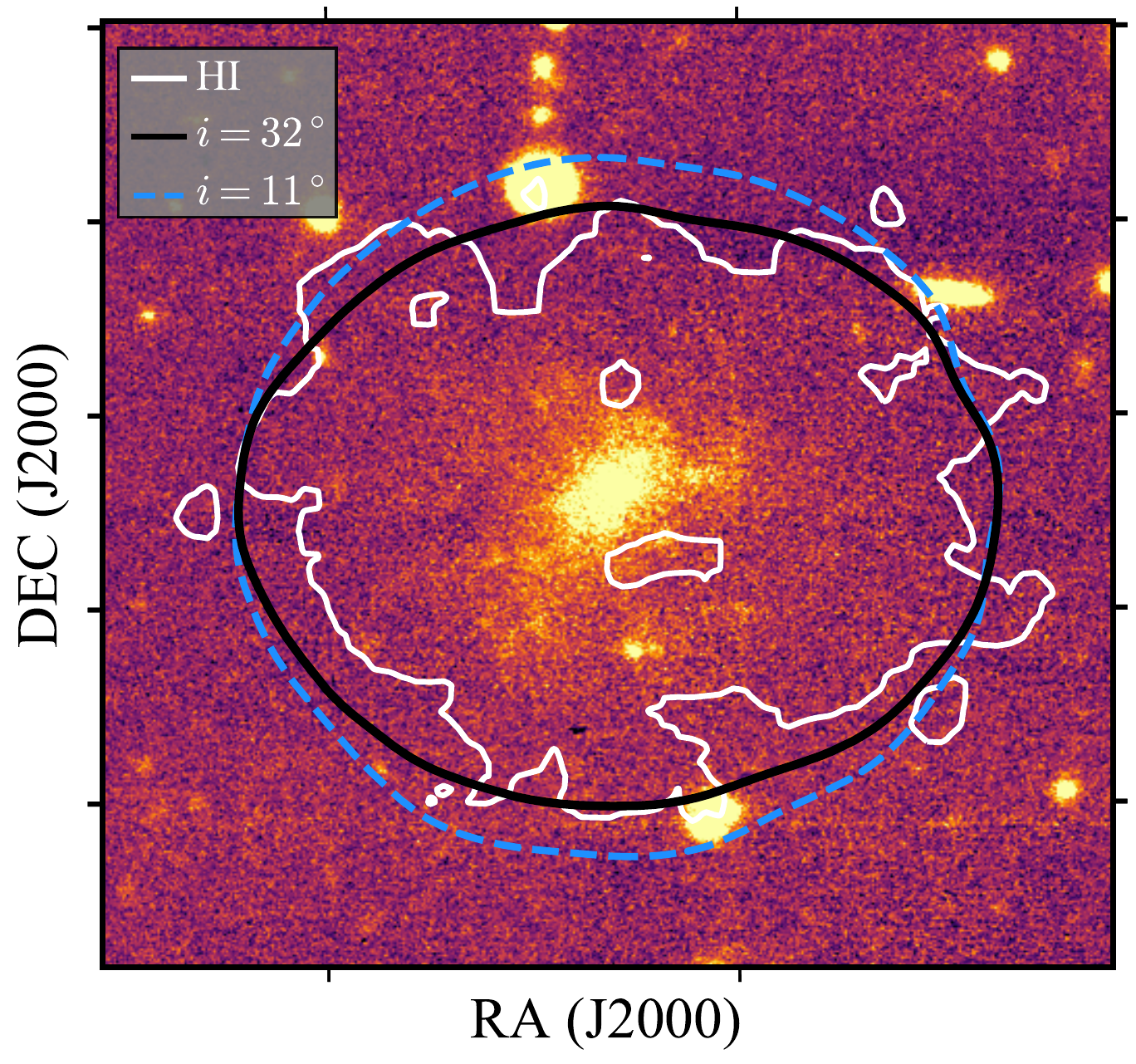}
    \caption{Comparison between the outer contours ($\textrm{S/N}=3$) of the H\,{\sc i} map of AGC~114905 (white) and two azimuthal models at different inclinations. While the model at $32^\circ$ (solid black line) provides a good fit to the data, the model at $10^\circ$ (dashed blue) is significantly more elongated than the data along the minor axis. The background shows the optical image of AGC~114905.}
    \label{fig:map_inclinations}
\end{figure}

The above values for the inclination are about $20^\circ$ degrees off the inclination we determined in Section~ \ref{sec:kinematics}. This is a discrepancy a factor 6--7 larger than the nominal uncertainty estimate of our measurement (see Fig.~\ref{fig:posterior_inc}), although inclinations below $\sim 25^\circ$ become increasingly difficult to constrain as ellipses with lower inclinations all have very similar shapes. We can also inspect visually if inclinations as low as $11^\circ-15^\circ$ can be consistent with the data. In Fig.~\ref{fig:map_inclinations} we show the outer contour of the H\,{\sc i} map of AGC~114905 overlaid on the $r-$band optical image, and we compare it with the equivalent contours of two $\mathrm{^{3D}}$Barolo azimuthal models (convolved with the observed beam) of AGC~114905 at different inclinations. The models are razor-thin axisymmetric discs (but see Section~ \ref{sec:inclination}). The model at $32^\circ$ does an overall good job at following the H\,{\sc i} contour, while the contour for the model at $11^\circ$ appears inconsistent with it, being significantly more elongated along the minor axis. This is also shown in Fig.~\ref{fig:channels} where we compare the channel maps of AGC~114905 with the channel maps of our best-fitting model and those of a model with an inclination of $11^\circ$.

Given all of the above, we find it unlikely that we are severely overestimating the inclination of our UDG, although this remains the largest source of uncertainty in our analysis. Something else to consider is that there are other gas-rich UDGs showing a similar set of properties, all at different inclinations (\citealt{huds2019,huds2020}, see also \citealt{sengupta2019,shi2021}, and the spatially unresolved data from e.g. \citealt{leisman2017,karunakaran2020}). This means that the inclinations of all of them would need to be overestimated by a large margin. Still, it is desirable to repeat our analysis with a gas-rich UDG at a similar resolution as we have now for AGC~114905, but at higher inclination, and we aim to do this in the near future. 

\section{Conclusions}
\label{sec:conclusions}
We obtained new H\,{\sc i} interferometric observations of the gas-rich ultra-diffuse galaxy (UDG) AGC~114905 using the Karl G. Jansky Very Large Array in its B-, C- and D-configurations. The new data, tracing the H\,{\sc i} emission up to 10~kpc from the galaxy centre, have a spatial resolution a factor about 2.5 higher than previous data, and confirm that AGC~114905 has a regularly rotating gas disc.

We performed 3D kinematic modelling of the data cube using $\mathrm{^{3D}}$Barolo, which allowed us to recover the intrinsic rotation curve and velocity dispersion profile of the galaxy. AGC~114905 has a regular rotation curve that reaches a flat part with a circular speed (after a minor correction for asymmetric drift) of about 23~km~s$^{-1}$. This result confirms that this UDG lies off the baryonic Tully-Fisher relation, as suggested by \citet{huds2019,huds2020} with low-resolution data.

The observed circular speed profile of our UDG can be explained almost entirely by the contribution of the baryons alone, with little room for dark matter within our observed outermost radius ($R_{\rm out} \approx 10$~kpc). Moreover, we found that the circular speed profile cannot be reproduced by standard cold dark matter (CDM) halos: the only possibility to find a good fit to the data is if the concentration of the halo is as low as $\sim 0.3$, completely off CDM expectations. We tested whether the rotation of our UDG is instead reproduced within the MOND framework, but we find that there is a significant mismatch on the normalization and shape of the MOND rotation curve with respect to our observations.

The geometry of the system (assumed to be an inclined axisymmetric disc) is the main source of uncertainty in our results. The inclination of AGC~114905 ($32 \pm 3^\circ$), which we measure from its total H\,{\sc i} map independently of its kinematics, is a significant caveat, but a number of independent pieces of evidence suggest that it cannot be overestimated to the extent of significantly changing the above results. Efforts to observe another gas-rich UDG at a similar spatial resolution but at higher inclination are under way. Finally, it is important to consider that we have confirmed for one UDG the robustness of the results obtained by \citet{huds2019,huds2020} at low resolution. The fact that the six UDGs (and see also e.g. \citealt{leisman2017,shi2021}) at different inclinations show the same behavior argues in favor of them being really exotic and suggests that our results are not the byproduct of systematic uncertainties.

We have strengthened and clarified previous results on the nature and startling dynamics of gas-rich UDGs. Yet, their origin and precise evolutionary pathways remain largely a mystery. The present work has also shown that gas-rich UDGs are a promising population to study dark matter, as they can potentially provide telltale clues to understand its nature.

\section*{Acknowledgements}
We would like to thank Federico Lelli and Justin Read for useful comments on this manuscript, as well as Alessandro Romeo for useful discussions on disc stability. We appreciate the feedback from an anonymous referee. P.E.M.P. and F.F. are supported by the Netherlands Research School for Astronomy (Nederlandse Onderzoekschool voor Astronomie, NOVA), Phase-5 research programme Network 1, Project 10.1.5.6. E.A.K.A. is supported by the WISE research programme, which is financed by the Netherlands Organization for Scientific Research (NWO). K.A.O. acknowledges support by the European Research Council (ERC) through Advanced Investigator grant to C.S. Frenk, DMIDAS (GA 786910).

This work is based on observations made with the Karl G. Jansky Very Large Array (VLA). The VLA is a facility of the National Radio Astronomy Observatory (NRAO). NRAO is a facility of the National Science Foundation operated under cooperative agreement by Associated Universities, Inc.

We have used extensively SIMBAD and ADS services, as well the Python packages NumPy \citep{numpy}, Matplotlib \citep{matplotlib}, SciPy \citep{scipy}, Astropy \citep{astropy}, spectral-cube \citep{spectral_cube}, and corner \citep{corner}, for which we are thankful.

\section*{Data Availability}
The datasets generated during and/or analysed during the current study are available from the corresponding author on reasonable request.



\bibliographystyle{mnras}
\bibliography{references} 




\appendix
\section{Channel maps}
\label{sec:channelmaps}

\begin{figure*}
    \centering
    \includegraphics[scale=0.45]{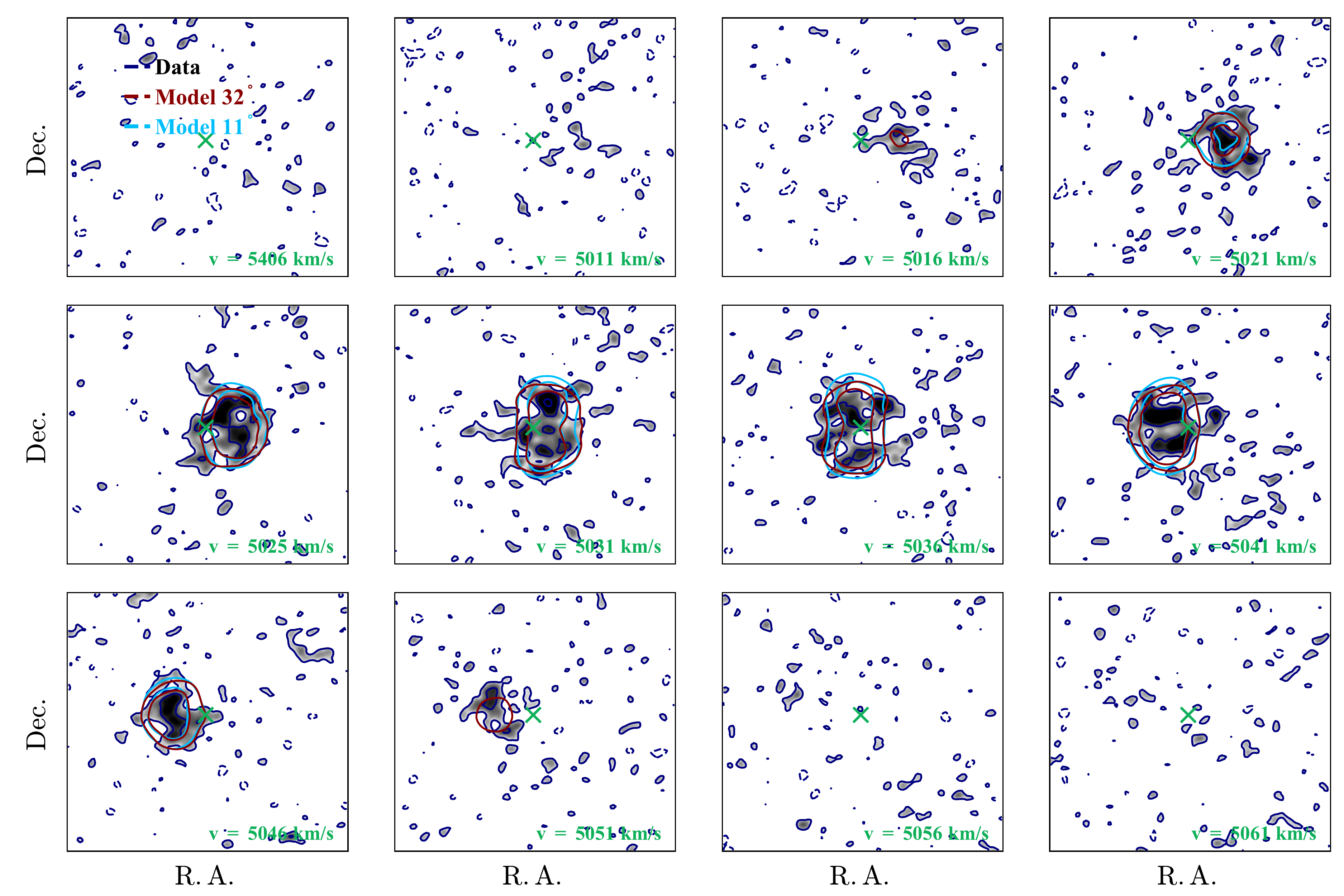}
    \caption{Representative channel maps of AGC~114905. The emission of the galaxy is shown in grey background and dark blue contours (open contours for negative values). The green crosses show the centre of the galaxy, and we indicate the velocity corresponding to each channel map on the bottom right corner. The contours for the best-fitting $\mathrm{^{3D}}$Barolo azimuthal tilted-ring model are shown in red, while the contours for a model at $11^\circ$ are shown in light blue. Contours are at -2, 2, 4 times the rms noise per channel.}
    \label{fig:channels}
\end{figure*}

Fig.~\ref{fig:channels} shows representative channel maps of the data cube of AGC~114905. The observed emission is shown in grey background and dark blue contours (open contours for negative values). The green cross shows the centre of the galaxy and the velocity of each channel map is given on the bottom right corner of each panel. In red, we show the contours for the best-fitting $\mathrm{^{3D}}$Barolo azimuthal tilted-ring model; while low S/N features are not fully reproduced, the model captures well the overall kinematics of the galaxy, as also shown in Fig.~\ref{fig:kinematics} with the PV diagrams. We also overlay in light blue the contours for a model with a fixed inclination of 11$^\circ$ (as needed to match CDM and MOND expectations, see Section~ \ref{sec:low_inc}). A close inspection shows that the model at $11^\circ$ has an excess of flux along the minor axis in the channels close to the systemic velocity; the model at $32^\circ$ does a better job in this regard (however, this comparison is better appreciated in Fig.~\ref{fig:map_inclinations}). Moreover, the model at $32^\circ$ matches in a better way the spectral extent of the observations.

\section{MCMC posterior distributions}
In this appendix we provide the main posterior distributions obtained with our MCMC analyses as described in the main text. 
\label{sec:appendix}
\begin{figure}
    \centering
    \includegraphics[scale=0.8]{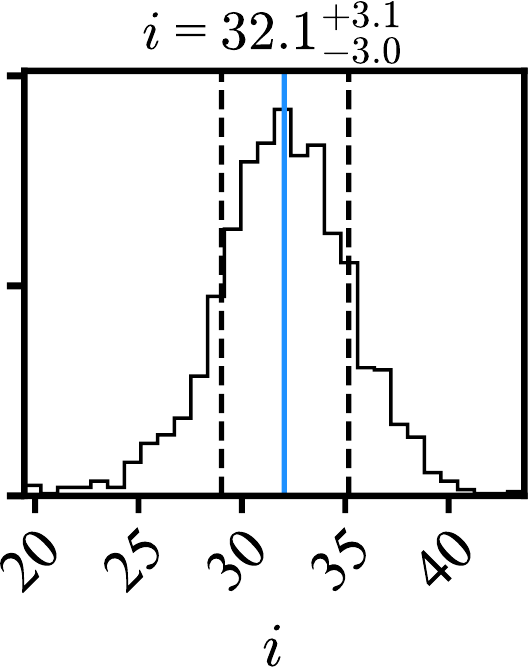}
    \caption{MCMC posterior distribution of the inclination of AGC~114905. The central value, shown in blue, is the median of the distribution, while the uncertainties represent the difference between the median and the $16^{\rm th}$ and $84^{\rm th}$ percentiles (dashed black lines). See Section~ \ref{sec:inclination} for details.}
    \label{fig:posterior_inc}
\end{figure}

\begin{figure}
    \centering
    \includegraphics[scale=0.36]{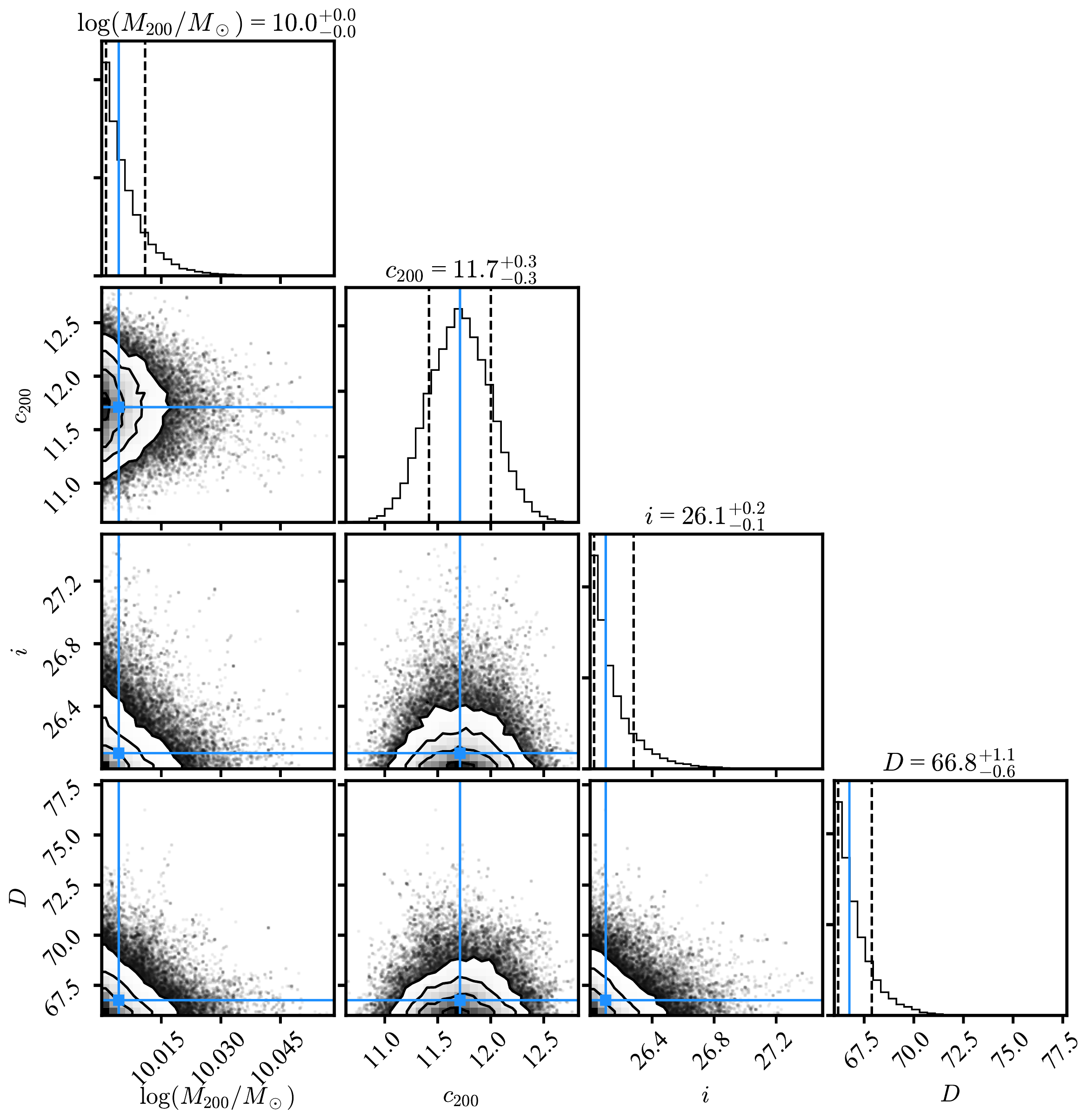}
    \caption{MCMC posterior distribution for our Case 1 mass model. Lines are as in Fig.~\ref{fig:posterior_inc}. See Section~ \ref{sec:massmodels} for details.}
    \label{fig:posterior_case1}
\end{figure}

\begin{figure}
    \centering
    \includegraphics[scale=0.36]{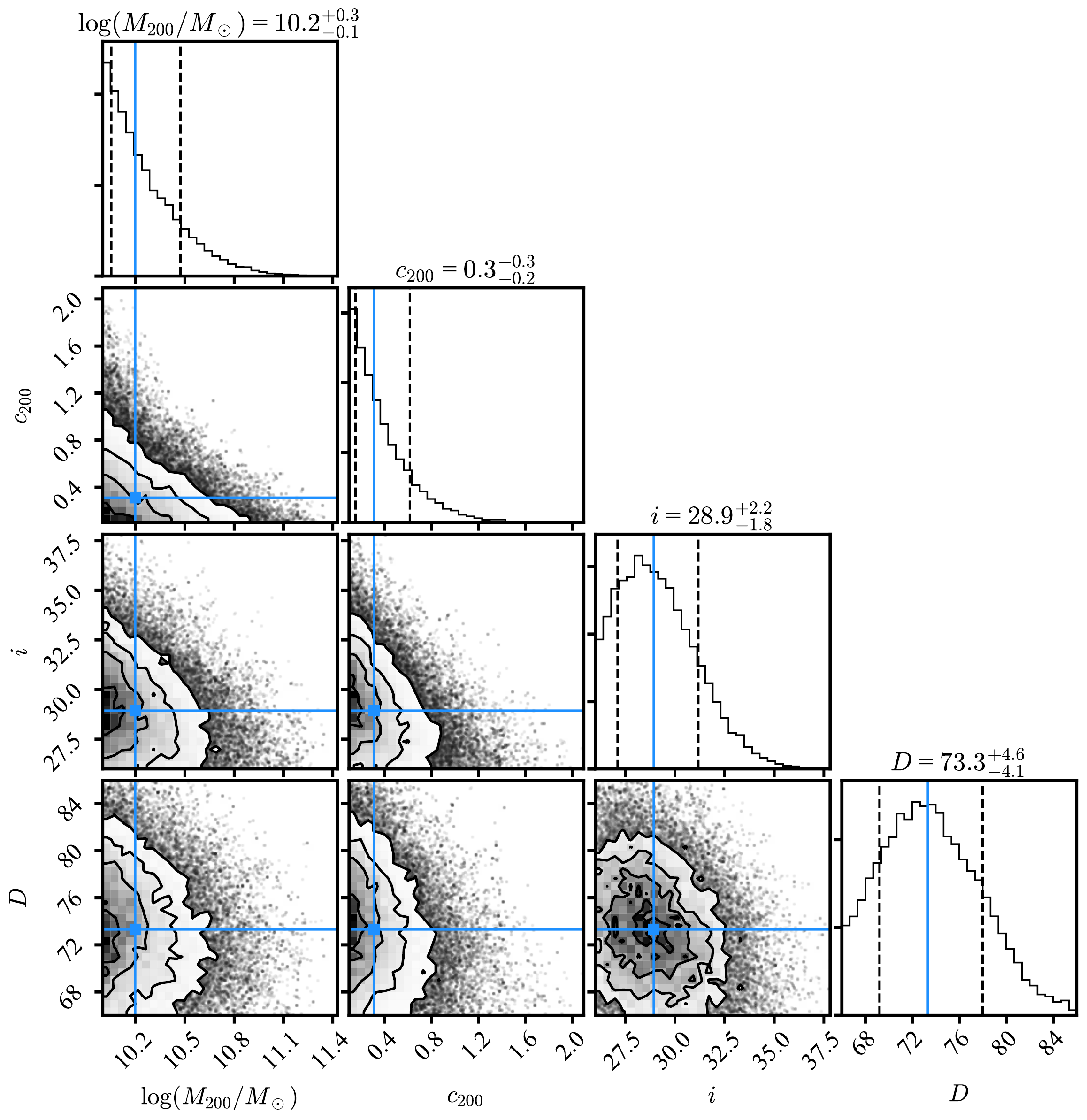}
    \caption{MCMC posterior distribution for our Case 2 mass model. Lines are as in Fig.~\ref{fig:posterior_inc}. See Section~ \ref{sec:massmodels} for details.}
    \label{fig:posterior_case2}
\end{figure}

\begin{figure}
    \centering
    \includegraphics[scale=0.6]{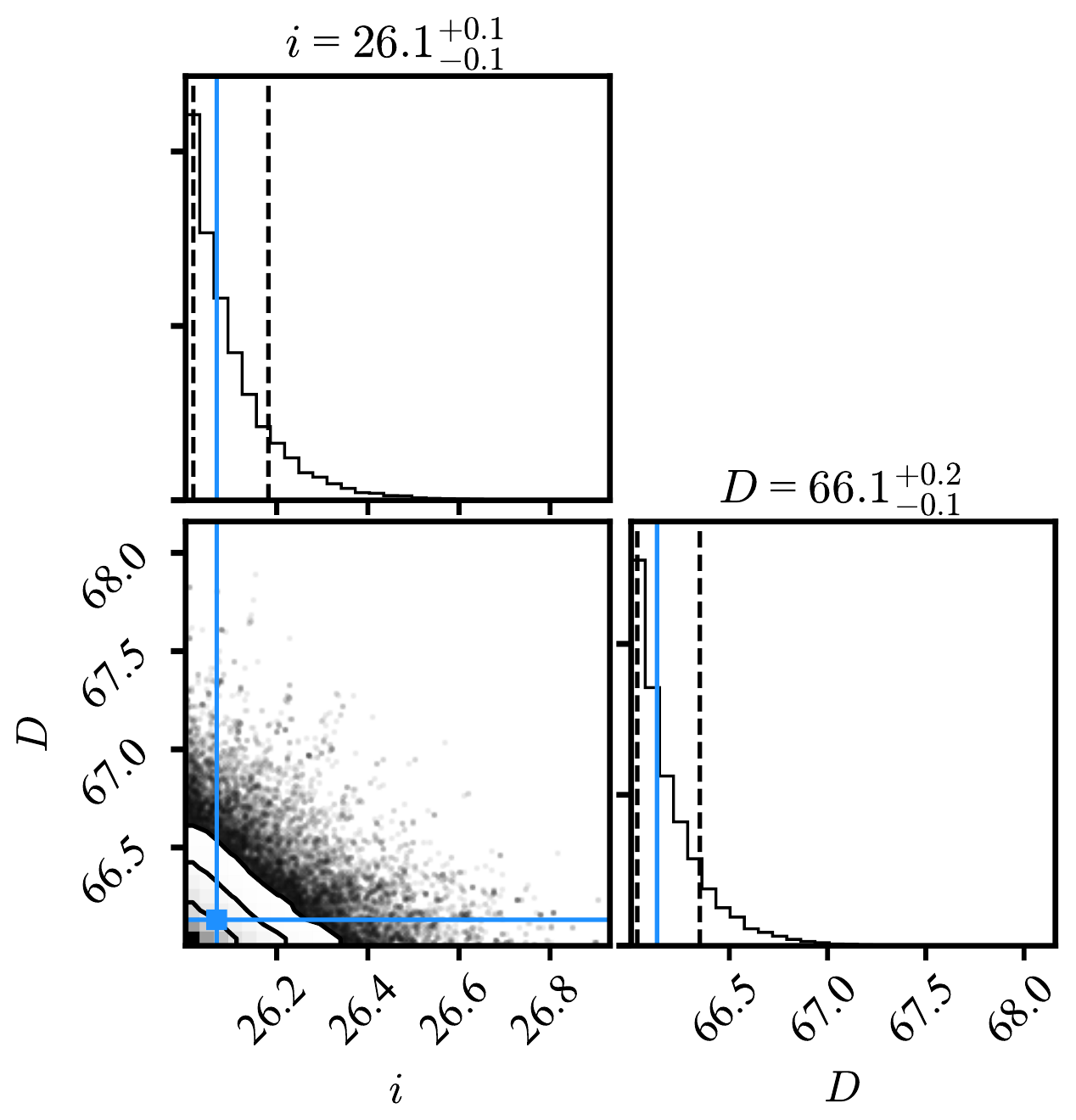}
    \caption{MCMC posterior distribution for the MOND model. Lines are as in Fig.~\ref{fig:posterior_inc}. See Section~ \ref{sec:mond} for details.}
    \label{fig:posterior_mond}
\end{figure}


\bsp	
\label{lastpage}
\end{document}